\documentclass[preprint]{elsarticle}

\usepackage{hyperref}
\journal{Earth and Planetary Science Letters}

\biboptions{authoryear}

\usepackage[inline]{trackchanges} 
\usepackage{soul}
\usepackage{amsmath}
\usepackage{tikz}
\usepackage{verbatim}
\usepackage{graphicx}
\usepackage{lineno}
\usepackage{setspace}
\usetikzlibrary{arrows,shapes}
\usepackage{natbib}
\usepackage{hyperref}
\usepackage{lineno}
\newcommand{\tidal}{\Pi}

\begin{document}

\begin{frontmatter}

\title{New insights into temperature-dependent ice properties and their effect on ice shell convection for icy ocean worlds}

\author[jackson,utig,oden]{Evan Carnahan}\corref{corr}
\cortext[corr]{Corresponding author}
\ead{evan.carnahan@utexas.edu}

\author[jackson,utig]{Natalie S. Wolfenbarger}

\author[rice]{Jacob S. Jordan}

\author[jackson,oden]{Marc A. Hesse}

\address[jackson]{Department of Geological Studies, Jackson School of Geosciences, The University of Texas at Austin, Austin, TX, USA}
\address[utig]{Institute for Geophysics, Jackson School of Geosciences, The University of Texas at Austin, Austin, TX, USA}
\address[oden]{Oden Institute for Computational Science and Engineering, The University of Texas at Austin, Austin, TX, USA}
\address[rice]{Department of Earth, Environmental and Planetary Sciences, Rice University, Houston, Texas, USA}
%
\begin{abstract}
\doublespacing
Ice shell dynamics are an important control on the habitability of icy ocean worlds. Here we present a systematic study evaluating the effect of temperature-dependent material properties on these dynamics. We review the published thermal conductivity data for ice, which demonstrates that the most commonly used conductivity model in planetary science represents a lower bound. We propose a new model for thermal conductivity that spans the temperature range relevant to the ice shells of ocean worlds. This increases the thermal conductivity at low temperatures near the surface by about a fifth. We show that such an increase in thermal conductivity near the cold surface can stabilizes the ice shell of Europa. Furthermore, we show that including temperature dependent specific heat capacity decreases the energy stored in the conductive lid which reduces the response timescale of the ice shell to thermal perturbations by approximately a third. This may help to explain surface features such as chaotic terrains that require large additions of energy to the near-surface ice.
\end{abstract}


\end{frontmatter}
\nolinenumbers
\doublespacing
\section{Introduction}
Several Jovian and Saturnian satellites are thought to host large subsurface oceans beneath their ice shells, maintained by strong tidal heating \citep[for review see][]{Nimmo2016}. These subsurface oceans represent prime candidates in the search for life beyond Earth \citep{Gaidos1999} and are targets for exploration by the upcoming JUICE and Europa Clipper missions \citep{Grasset2013,Pappalardo2015}. These missions aim to investigate the icy moons of Jupiter to better constrain conditions that may govern their habitability. This includes ice shell properties such as chemistry and thickness which are controlled by the dynamics of the ice shell. Convection in the ice shell could facilitate the delivery of radiolytically produced oxidants from the surface to the ocean, enhancing the potential for life within a reduced ocean \citep{Vance2016}. This study focuses on constraining the convective dynamics in the Europan ice shell, but our conclusions apply to other similar bodies. 

Multiple studies have investigated whether the Europan ice shell experiences convective overturn \citep[for example][]{Ojakangas1989,Pappalardo1998,McKinnon1999,Kattenhorn2014,Howell2018}. Convection determines ice shell thickness, the rate of heat loss and the material transport across the ice. The convective stability of the ice shell is governed by the material properties of ice (viscosity, density, heat capacity, and thermal conductivity), internal heating caused by tidal dissipation, and surface temperature. Because maximum tidal dissipation occurs at a viscosity near the melting temperature of ice, these ice shells are highly susceptible to partial melting in the convecting portion of the ice shell \citep{Sotin2002,Tobie2003,Kalousova2017,Vilella2020}. 

\begin{figure}
    \centering
    \includegraphics[width = \linewidth]{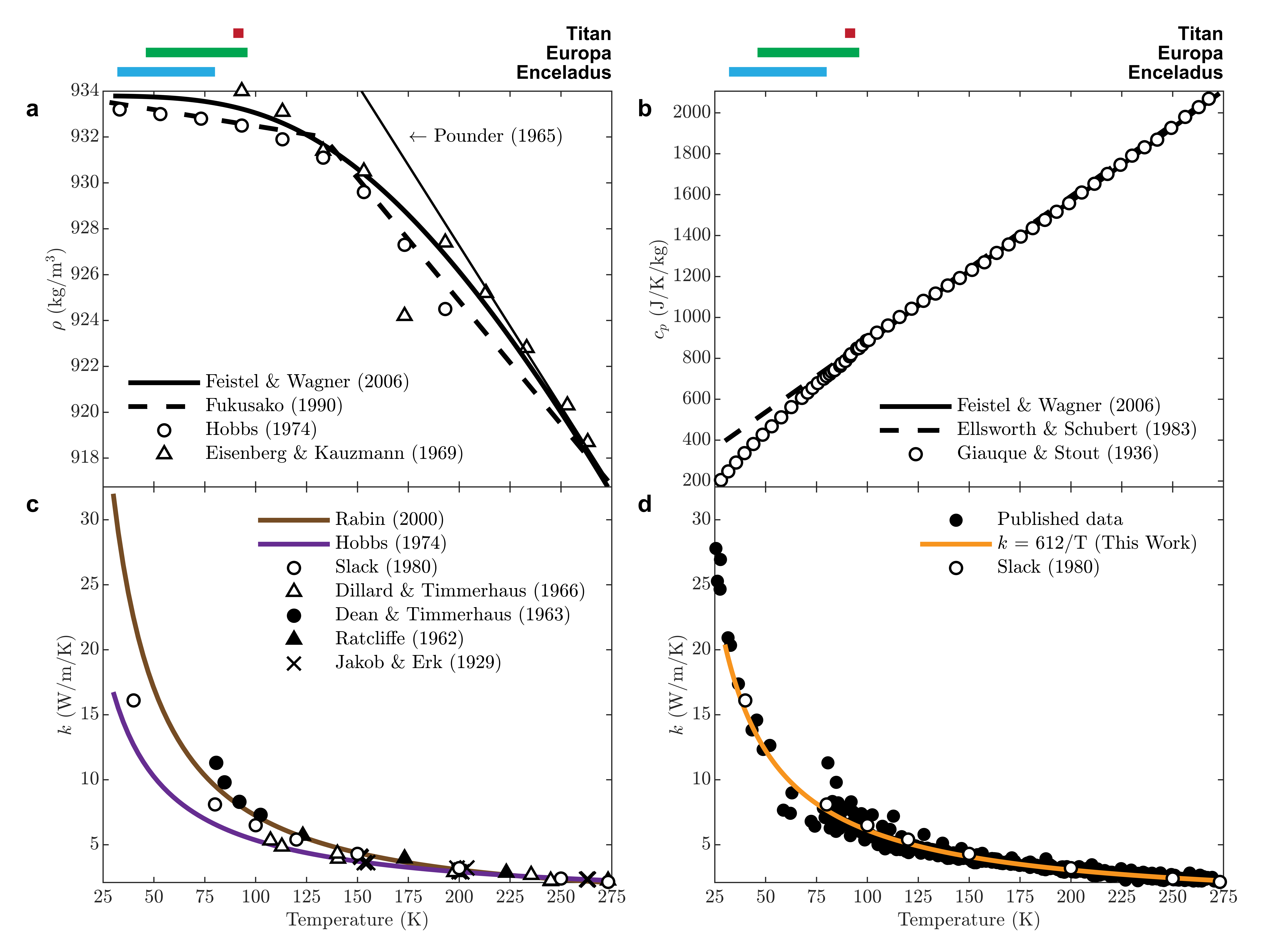}
     \caption{Temperature dependent ice properties: (a) models of density compared to experimental data points (b) models of isobaric specific heat capacity compared to the dataset of \citet{Giauque1936} (c) thermal conductivity models of \cite{Hobbs1974} and \cite{Rabin2000} and select values from the datasets used to obtain them (d) our model of thermal conductivity and the collection of published thermal conductivity data used to obtain the fit compared to the "best estimate" values of \citet{Slack1980}. The range of mean annual surface temperatures at Titan \citep{Jennings2016}, Europa \citep{Ashkenazy2019}, and Enceladus \citep{Weller2019} for reference.} 
    \label{fig:iceProp}
\end{figure}

The material properties of ice are highly temperature-dependent. For icy ocean worlds the temperatures in the ice shell can range from 30 K at the surface to 273 K near the ice-ocean interface. This leads to large variations in material properties across these ice shells and strongly affects their dynamics. The relatively small change in ice density, shown in Figure~\ref{fig:iceProp}a, provides the driving force for convection but is typically linearized in numerical simulations. Most previous simulations have considered the Arrhenius dependence of viscosity on temperature (not shown) that leads to an exponential increase with declining temperature \citep{Mitri2005,Showman2005,AlluPeddinti2015,Howell2018,Peddinti2019,Weller2019}.
This gives rise to extreme asymmetry in the conductive boundary layers \citep{Solomatov1995}, causing the ice shell to be divided into a cold, brittle conductive lid overlying warm, ductile convecting ice. In contrast, few simulation studies have considered the temperature dependence of the other material properties. The specific heat of ice near the surface is only a quarter of its value at the base (Figure~\ref{fig:iceProp}b), yet this change is neglected in previous work on ice shell convection. Below we focus on the variation in thermal conductivity with temperature (Figure~\ref{fig:iceProp}c). 

Although thermal conductivity is an important parameter governing the convective stability of an ice shell, the experimental data show considerable variance over the range of temperatures relevant to icy ocean worlds. 
Two common conductivity models, \cite{Hobbs1974} and \cite{Rabin2000}, are shown in Figure~\ref{fig:iceProp}c. They represent lower and upper bounds of the published experimental data for thermal conductivity of ice Ih, respectively. A limitation of the Hobbs and Rabin models is that they are highly tuned to the datasets from which they were derived. In particular we note that the Hobbs model included the low thermal conductivity data from \citet{Dillard1966}, which was later deemed to be inaccurate \citep{Slack1980}. In applications to planetary sciences, where temperature-dependence is taken into account, the conductivity of ice Ih is generally assumed to follow the Hobbs model \cite[e.g.,][]{McKinnon1999,Tobie2003,Hammond2016,Kalousova2017,Hesse2019} which may under predict the thermal conductivity by approximately a fifth. Works that assume a constant thermal conductivity (e.g. \citet{Hussmann2004,Mitri2005,Barr2005,Robuchon2011,Hammond2014,Weller2019,Vilella2020}) could underpredict the thermal conductivity by an order of magnitude at the surface of the ice shell. 

To obtain a model that best represents the thermal conductivity of ice Ih, we fit a collection of the available published data that spans the temperatures relevant to icy ocean worlds (Figure~\ref{fig:iceProp}d). We choose to model the thermal conductivity as inversely proportional to temperature, motivated by the simple theoretical model applicable to high temperature, monocrystalline ice at constant volume \citep{Andersson1980}. This model represents the most comprehensive, simplest, and best fit of available experimental data to date and serves as an intermediate model between Hobbs and Rabin (Figure~\ref{fig:iceProp}c and d). Details of the experimental datasets and the new model proposed here are given in the supplementary material. It is important to acknowledge that thermal conductivity data are highly sensitive to sample characteristics and preparation, such as anisotropy \citep{Klinger1975} and freezing rate \citep{Bonales2017}. Generally experiments are quoted to have accuracy on the order of 10\%, which is consistent with the deviation we observe around our fit \citep[see][and references therein]{Slack1980}.

Here we present a systematic study of the hydrodynamic stability of ice shells with temperature-dependent material properties. This is motivated by the early theoretical work of \citet{McKinnon1999} and more recent simulations by \citet{Kalousova2017}, which demonstrates that even the lower bound on the temperature-dependence of thermal conductivity can suppress convection. We are further motivated by the large variance in the aforementioned data for the thermal conductivity of ice at low temperature. We adopt three models for the temperature dependence of thermal conductivity to span the range of uncertainty in experimental data: the lower bound given by \cite{Hobbs1974}, the upper bound given by \cite{Rabin2000}, and a new intermediate model derived here. While the focus of this contribution is primarily on the effect of thermal conductivity variations on ice shell stability, we also investigate the effect of temperature-dependent specific heat on ice shell response time. 
Here we use the linear model for specific heat introduced by \citet{Ellsworth1983}. We note that we found the effect of the non-linear density variation at low temperature (Figure~\ref{fig:iceProp}a) to be minor and will hence not discuss it further.

\section{Ice shell convection with temperature-dependent properties}
The standard model for ice shell convection arises from the balance of mass, energy, and momentum for an incompressible Newtonian fluid under the Oberbeck-Boussinesq approximation. The model problem is described by the following system of three governing equations,

\begin{eqnarray} 
    \label{eqn:stokesLM}
    \nabla \cdot [\eta (\nabla \mathbf{u}+\nabla \mathbf{u}^T)]-\nabla p = \rho g \hat{\mathbf{z}},\\
    \label{eqn:stokesCOM}
    \nabla \cdot \mathbf{u} = 0,\\
    \rho c_p \frac{\partial T}{\partial t} +\nabla \cdot \left[\mathbf{u} \rho c_p T - k \nabla T \right] = G. \label{eqn:COT}
\end{eqnarray}
Here, the unknowns are the velocity vector $\mathbf{u}$, pressure $p$, and temperature $T$. The gravitational acceleration, $g$, is assumed to be constant across the ice shell. All material properties of the ice are allowed to vary with temperature (Figure~\ref{fig:iceProp}): dynamic viscosity, $\eta=\eta(T)$; density, $\rho=\rho(T)$; specific heat capacity, $c_p = c_p(T)$ and thermal conductivity, $k=k(T)$. All properties except viscosity are shown in Figure~\ref{fig:iceProp} and parameterizations for all material properties are given in the supplementary materials. We note that changes in density are only retained in the right hand side of \eqref{eqn:stokesLM}, due to the Oberbeck-Boussinesq approximation. We use the standard tidal heating term, $G$, and assume that the maximum dissipation occurs at the melting point viscosity \citep{Tobie2003}, see supplementary materials for details. 
Here we assume a Newtonian rheology with an Arrhenius dependence on temperature, similar to most previous work (e.g. \citet{Tobie2003,Hussmann2004,Mitri2005,Kalousova2017,Weller2019,Vilella2020}), but acknowledge that non-Newtonian rheology has been shown to affect convective stability \citep{Barr2005}.

The governing equations (\ref{eqn:stokesLM}-\ref{eqn:COT}) are solved on a rectangular domain of thickness, $d$, and width, $w$, with free-slip boundary conditions for the velocity. At the top of the domain we impose a constant surface temperature, $T_s$, and the basal temperature, $T_b$, is set to the low pressure melting point of ice, $T_m=273$ K. The temperature field is initialized with a perturbed steady conductive profile, see Section~\ref{sec:results}. All simulations use the nonlinear density dependence shown in Figure~\ref{fig:iceProp}a and the linear specific heat relation shown in Figure~\ref{fig:iceProp}b, unless otherwise noted.

To solve \eqref{eqn:COT} with temperature-dependent specific heat capacity we introduce the enthalpy of the ice, given by
\begin{align} \label{eqn:totalEnthalpy}
H(T)=\rho_0 \left[ h_0 + \int_{T_0}^T c_p(\tau) d\tau \right],
\end{align}
where $h_0$ is the reference enthalpy at temperature $T_0$ and $c_p$ is the variable specific heat. Here we use the melting point of pure ice as the reference temperature for enthalpy, so that $T_0 = T_b = T_m$ and $h_0(T_0)=0$. We note that due to the Oberbeck-Boussinesq approximation, density has been assumed constant here, $\rho_0=\rho(T_0)$, but variations in $\rho$ can be included. The conservation of energy is then written,
\begin{align}\label{eqn:COE}
    \frac{\partial H}{\partial t} +\nabla \cdot\left[\mathbf{u} H - \kappa \nabla H \right] = G,
\end{align}
where $\kappa$ is the thermal diffusivity, defined as $\kappa = k \frac{dT}{dH}$. The additional term, $\frac{dT}{dH}$, is the inverse of the volumetric heat capacity, which arises when the chain rule is applied to the temperature gradient. 

For a constant specific heat and density, $H = \rho c_p (T-T_0)$, so $\frac{dT}{dH} = \frac{1}{\rho c_p}$, and \eqref{eqn:COE} simplifies to \eqref{eqn:COT}. The algebraic expressions for $\frac{dT}{dH}$ with temperature-dependent specific heat capacity are given in the supplementary information. The formulation of the energy conservation equation in terms of enthalpy is commonly referred to as the Enthalpy Method \citep[see][]{Alexiades1993}. This method can be extended to account for phase change and is therefore of broader interest to models of ice shell convection that include partial melting.

The non-dimensionalization of \eqref{eqn:stokesLM}, \eqref{eqn:stokesCOM} and \eqref{eqn:COE} is given in the supplementary materials and results in the following three independent dimensionless parameters
\begin{linenomath*}
\begin{align}
    \label{eqn:nonParam}
    \textrm{Ra} = \frac{g  \Delta \rho\, d^3}{\eta_c \kappa_{c}}, \quad 
    \tidal = \frac{G_{\text{max}}\, d^2}{k_c \Delta T}, \quad
    \mathrm{and}
    \quad \Theta = \frac{T_s}{T_b}.
\end{align}
Here, $d$ is the ice shell thickness, $\Delta T = T_b - T_s$ and $\Delta \rho=\rho(T_b) - \rho(T_s)$ are the temperature and density change across the ice shell, and $G_{\textrm{max}}$ is the maximum tidal heating. The characteristic material properties $\eta_c$, $k_c$, and $\kappa_{c}$ are defined at the melting point of ice, see supplementary information. 
\end{linenomath*}

The first parameter is the basal Rayleigh number, $\textrm{Ra}$, which is the ratio between the characteristic times for heat conduction and advection across the ice shell. The second is the tidal heating number, $\tidal$, which measures the ratio of tidal heat production to conductive heat transport. The third is the homologous temperature of ice at the surface, $\Theta$, which is introduced by the temperature dependence of material properties and governs the contrast in material properties across the ice shell. A fourth geometric dimensionless parameter is the aspect ratio of the domain, $Ar=w/d$. This parameter is less important, because we are interested in laterally extensive domains.

\begin{figure}
    \centering
    \includegraphics[width = \textwidth]{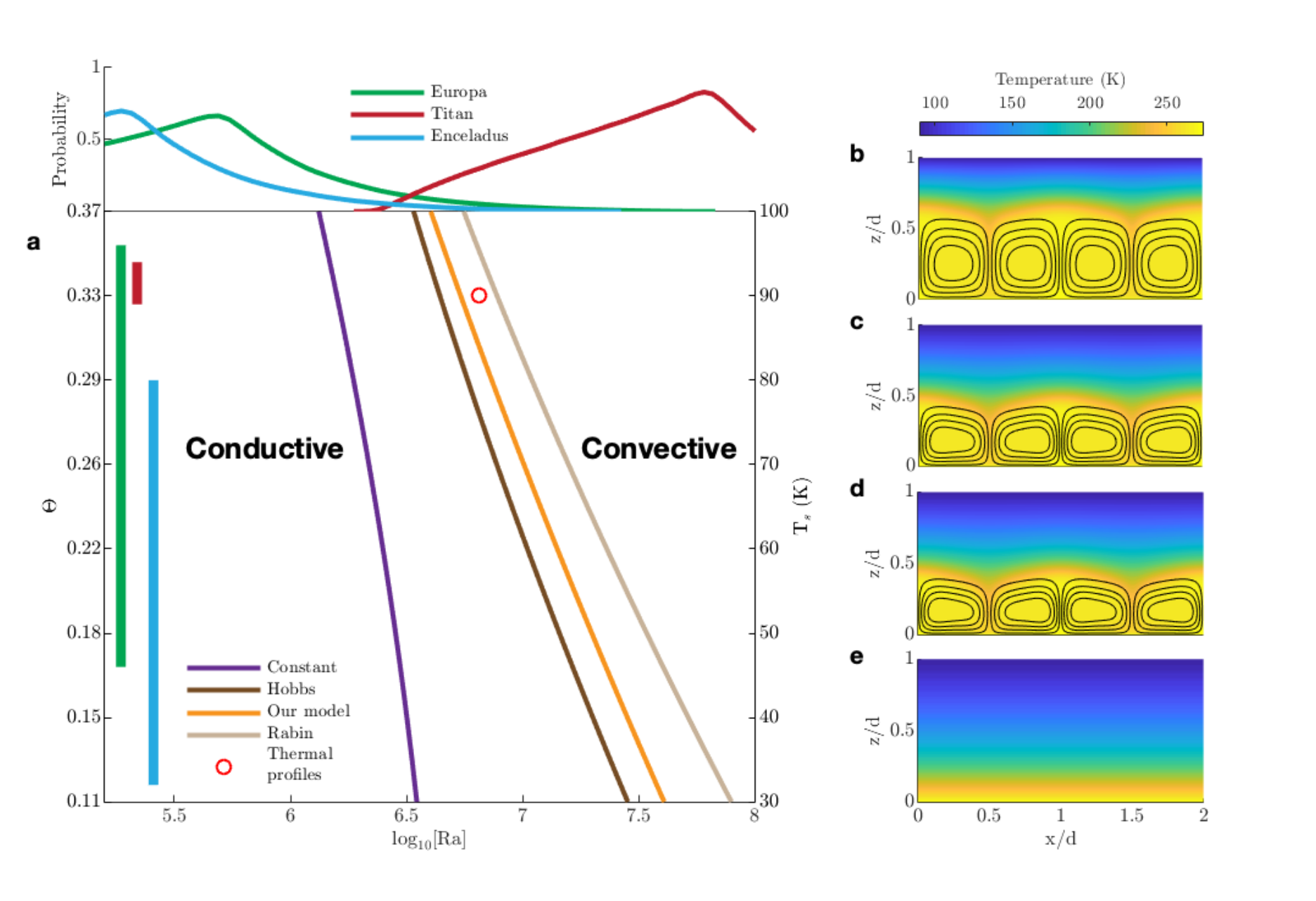}
    \caption{Ice shell stability with temperature-dependent thermal conductivity ($\tidal=0$): (a) Regime diagram with the stability boundary in Ra$\Theta$-space. Parameter ranges relevant to Enceladus, Europa and Titan are indicated. The red circle gives the parameter combination (Ra = $6.5\cdot10^{6}$, $\theta=0.33$/$T_s=90 \ \mathrm{K}$) for temperature fields and streamlines shown in panels (b-e) with four different $k(T)$ relations: (b) constant: $k=k(T_b)=2.26\ \mathrm{W} \ \mathrm{m}^{-1}\ \mathrm{K}^{-1}$, (c) \cite{Hobbs1974}, (d) this work, (e) \cite{Rabin2000}.}
    \label{fig:stabVarKappa}
\end{figure}

The parameterizations for the temperature dependence of specific heat, thermal conductivity, and the exponent in the Arrhenius relation for viscosity can all be written in the functional form, $f = a+b T^v$. The dimensionless form of the temperature dependence is given by
\begin{align}\label{eqn:kappa non-dim}
    f'(T',\Theta) = \pi_1 + \pi_2 ((1-\Theta)T'+\Theta)^{\pi_3},
\end{align}
where $T'=(T-T_s)/\Delta T$ is the dimensionless temperature. The three dimensionless parameters $\pi_1=\frac{a}{a+bT_b^v}$, $\pi_2=\frac{bT_b^v}{a+bT_b^v}$, and $\pi_3=v$ are constant as long as the basal temperature, $T_b$, does not change. 
Therefore, the only dimensionless group introduced by the temperature dependence is the homologous temperature at the surface, $\Theta$. 

The governing equations \eqref{eqn:stokesLM}, \eqref{eqn:stokesCOM}, and \eqref{eqn:COE} are solved for $\mathbf{u}$, $p$ and $H$ on a staggered Cartesian mesh with conservative finite differences for the Stokes equation and a finite volume method with flux-limiters and an explicit adaptive time-stepping scheme for the energy equation \citep{LeVeque1992}. Streamlines are calculated as equally spaced contours of the streamfunction. Details of the numerical implementation and standard benchmark tests are given in the supplementary materials. All simulations reported here use a domain with aspect ratio, $Ar = 2$, and a 120 by 120 tensor product mesh. 

\section{Simulation Results}\label{sec:results}
We present a systematic study of the effect of the three dimensionless parameters, Ra, $\tidal$, and $\Theta$ on the hydrodynamic stability of a generic ice shell. The initial conditions for all simulations are steady state conductive geotherms. The steady state profile is perturbed by a small amplitude, mode two sinusoidal temperature variation (smaller than $0.175$ K). We calculate potential Ra combinations for Europa, Titan, and Enceladus by uniformly sampling the likely uncertainty in the parameter ranges for bottom viscosity, $10^{13} - 10^{15}$ \citep{Tobie2003}, ice shell thickness \citep{Vance2018}, and surface temperature \citep{Jennings2016,Ashkenazy2019,Weller2019}. Furthermore, for Europa, we estimate the tidal parameter, $\tidal$, from thickness, tidal heating rate, and surface temperature \citep{Tobie2003}.

First we study the hydrodynamic stability of ice shells in the absence of tidal heating, $\tidal=0$, to focus on the effect of the thermal conductivity relations, $k(T)$, shown in Figures~\ref{fig:iceProp}c and \ref{fig:iceProp}d. The stability boundaries for different conductivities in Ra$\Theta$-space and the likely parameter ranges for Enceladus, Europa and Titan are shown in Figure~\ref{fig:stabVarKappa}a. As expected, the hydrodynamic stability is most sensitive to Ra and we can view the stability boundary as a $\Theta$-dependent critical Rayleigh number, Ra$_c(\Theta)$. For constant thermal conductivity values, Ra$_c$ is between $1.2\cdot10^{6}$ and $3.4\cdot10^{6}$, comparable to previous work \citep{McKinnon1999,Barr2005}. The critical Ra number increases with decreasing homologous surface temperature, $\Theta$. This stabilization of the ice shell is most pronounced for the upper-bound Rabin model. 

The choice of $k(T)$ relation can shift Ra$_c(\Theta)$ for convection by more than one order of magnitude at low $\Theta$. At low $\Theta$ uncertainty in the $k(T)$ relation results in almost a factor of three change in Ra$_c$, from $10^{7.45}$ to $10^{7.9}$. The effect of different $k(T)$ relations on ice shell dynamics are illustrated in Figures~\ref{fig:stabVarKappa}b to \ref{fig:stabVarKappa}e which show solutions for a parameter combination appropriate for Europa's ice shell (red circle in Figure~\ref{fig:stabVarKappa}a). The solution assuming constant $k$ in Figure~\ref{fig:stabVarKappa}b shows the expected stagnant conductive lid underlain by a thick convecting region \citep{Tobie2003,Showman2004,Mitri2005}. Introducing even the lower bound \citet{Hobbs1974} relation for $k(T)$ almost doubles the thickness of the stagnant conductive lid \citep{Tobie2003,Kalousova2017}. Choosing the upper bound \citet{Rabin2000} relation for $k(T)$ stabilizes the ice shell and leads to a conductive geotherm. For the intermediate $k(T)$ relation advocated here results are closer to the Hobbs model than the Rabin model.

\begin{figure}
    \centering
    \includegraphics[width = \textwidth]{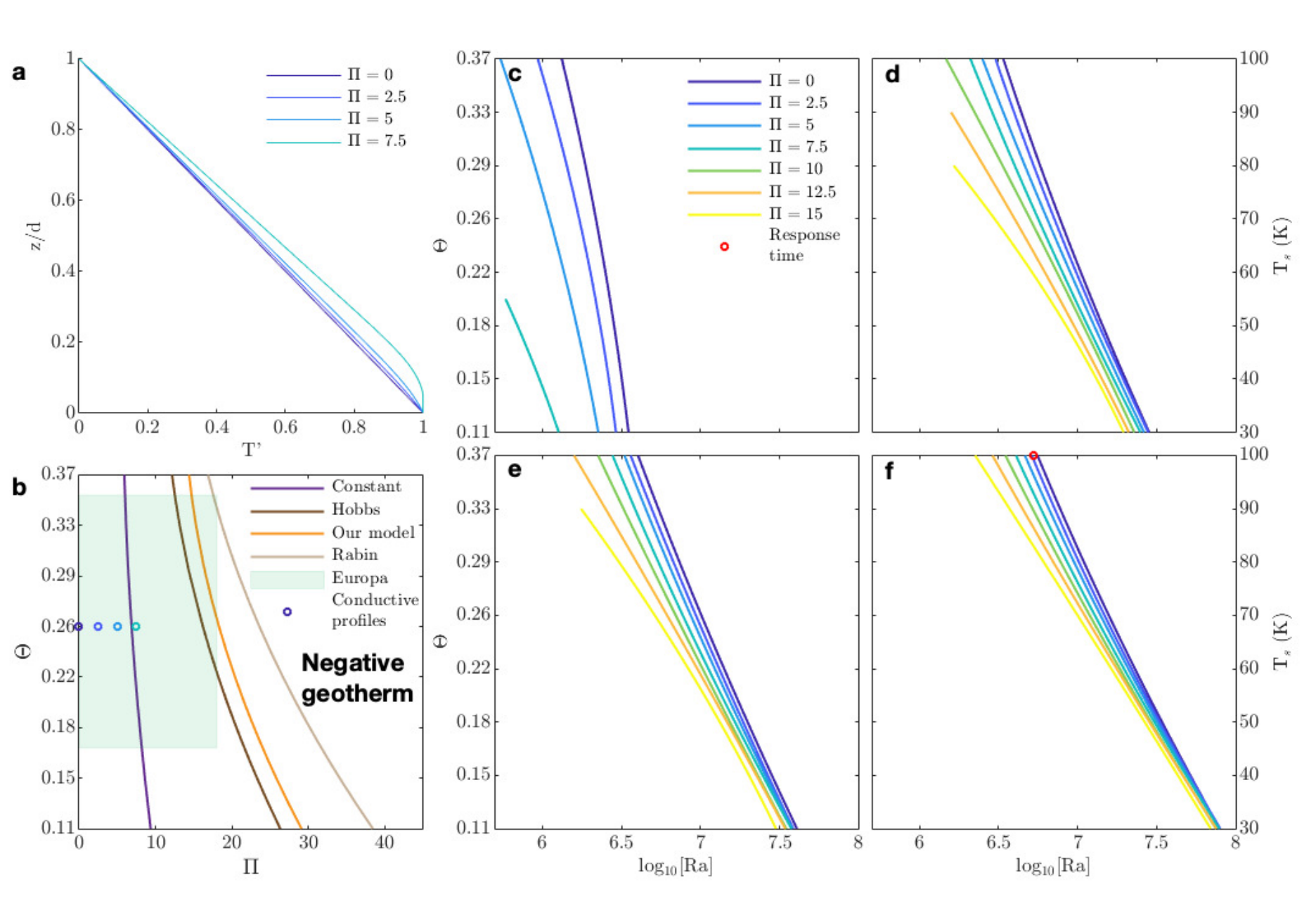}
    \caption{Ice shell stability with tidal heating: (a) Conductive profiles for constant $k$ with increasing tidal heating, $\tidal$. (b) Feasible regions in $\tidal\Theta$-space that result in conductive geotherms with non-negative basal heat flow for different $k$ relationships. (c-f) Regime diagrams in Ra$\Theta$-space that show the effect of tidal heating on the stability boundary for different $k$ relationships: (c) constant: $k=k(T_b)=2.26 \ \mathrm{W} \ \mathrm{m}^{-1}\ \mathrm{K}^{-1}$, (d) \cite{Hobbs1974}, (e) this work, (f) \cite{Rabin2000} with parameter combination used for response time to tidal perturbation, Figure \ref{fig:cpRespTime}, marked.}
    \label{fig:stabVarPiTide}
\end{figure}

Next, we consider the effect of tidal heating on the stability boundary, Ra$_c(\Theta,\tidal)$. For the fixed temperature boundary conditions considered here, tidal heating can produce conductive geotherms with a temperature maximum in the ice shell resulting in downward heat flow from the ice into the underlying ocean (Figure~\ref{fig:stabVarPiTide}a). To avoid unphysical initial conditions, which have previously been simulated, e.g.- \citet{Kalousova2017,Vilella2020}, we restrict our study of hydrodynamic stability to values of $\tidal$ that lead to non-negative basal heat flow (Figure \ref{fig:stabVarPiTide}b). Temperature-dependent thermal conductivity increases require larger tidal heating to produce downward heat flow. For Europa, this results in approximately a doubling of the possible starting conductive profiles in $\tidal \Theta$ parameter space (Figure \ref{fig:stabVarPiTide}b). The effect of $\tidal$ on the stability boundary in Ra$\Theta$-space is shown in Figures~\ref{fig:stabVarPiTide}c-f. Increasing $\tidal$ decreases Ra$_c$ for all $k(T)$ relations and this destabilization of the ice shell is strongest at high $\Theta$. Comparing Figure~\ref{fig:stabVarPiTide}c with Figure~\ref{fig:stabVarPiTide}e shows that tidal heating has a smaller effect on the stability of an ice shell than the choice of the thermal conductivity relation at all surface temperatures.

\begin{figure}
    \centering
    \includegraphics[width = \textwidth]{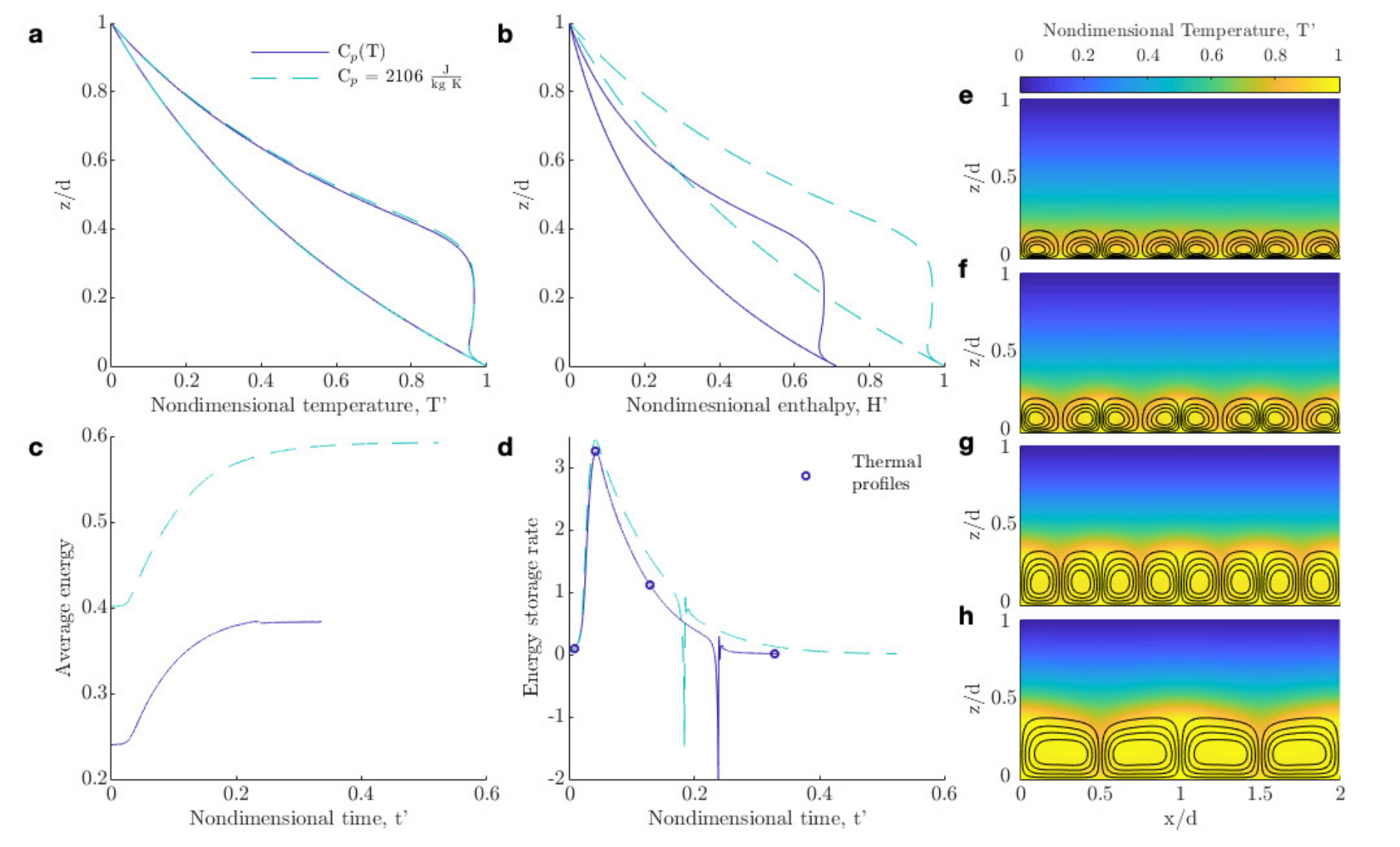}
    \caption{Ice shell response from an initially conductive ice shell with $\tidal=0$, to a convecting ice shell after an increase in tidal heating to $\tidal=2.5$. Results shown for both $c_p(T)$ \citep{Ellsworth1983} and $c_p = c_p(T_b) = 2106 \  \mathrm{J}\  \mathrm{kg}^{-1}\  \mathrm{K}^{-1}$. Ra$ =5.3 \cdot 10^6$, $\theta=0.37$, and \cite{Rabin2000}. Enthalpies are negative according to \eqref{eqn:totalEnthalpy}, but have been shifted so they are zero at the surface to improve readability. Horizontally averaged profiles of (a) dimensionless temperature and (b) dimensionless enthalpy. (c) Average dimensionless energy in the ice shell during the transition from conductive to convective steady states. (d) Dimensionless rate of energy into the ice shell, merging of convective cells results in a momentary decrease in energy in the ice shell. (e-h) Snapshots of temperature in the ice shell throughout the onset of convection for $c_p(T)$.}
    \label{fig:cpRespTime}
\end{figure}

Finally, we consider the temperature dependence of the specific heat capacity of ice, $c_p(T)$, shown in Figure~\ref{fig:iceProp}b. The heat capacity does not affect steady conductive or quasi-steady convective geotherms (Figure~\ref{fig:cpRespTime}a). As such, it does not influence the stability of the ice shell, but it determines the response timescale to thermal perturbations. Here we consider the response of a conductive ice shell (Ra$ = 5.3 \cdot 10^6$, $\Theta = 0.37$) to an increase in tidal heating from $\tidal = 0$ to $\tidal = 2.5$ that induces convection. The energy stored in the cold conductive portion of the ice shell is significantly less for temperature-dependent heat capacity (Figure \ref{fig:cpRespTime}b and c). As such, the $c_p(T)$ model requires less energy input from tidal heating to transition from steady conductive to quasi-steady convective (Figures \ref{fig:cpRespTime}c and d). Therefore, the response timescale of the model with temperature-dependent heat capacity is approximately 20 to 40\% faster (Figures \ref{fig:cpRespTime}c and d). The change in response timescales also affects the timing of convective dynamics, for example the merging of convection cells, Figures \ref{fig:cpRespTime}d, g, and h. Which further impacts the growth rate of the ice shell, \cite{Peddinti2019}.

\section{Discussion} 
The habitability of the internal oceans of icy ocean worlds is highly dependent on the poorly constrained properties of the outer icy shell. Future missions to explore and study these ocean worlds require bounds on ice shell properties, which govern design drivers for instruments such as ice shell probes and ice penetrating radar. Although some important properties, such as the grain size of the ice, cannot be determined \textit{a priori}, the temperature dependence of most material properties of ice Ih are relatively well constrained from laboratory studies (Figure~\ref{fig:iceProp}). Although virtually all numerical simulations of ice shell convection consider the temperature dependence of viscosity, the temperature dependence of other material properties has largely been neglected.

Whereas data for density and specific heat capacity of ice Ih are relatively unambiguous, the available experimental data for the thermal conductivity of ice Ih show increasing variation with decreasing temperatures. Existing parameterizations typically fit a subset of these data and can differ by a factor of two when they are extrapolated to temperatures on the surface of icy ocean worlds. This uncertainty results in up to a factor of three change in Ra$_c$. Therefore, which thermal conductivity relation is chosen may determine if the ice shell convects or not (Figure~\ref{fig:stabVarKappa}b-\ref{fig:stabVarKappa}e). This is particularly relevant for Europa, where most Ra numbers lie on the stable side of the stability boundary (Figure~\ref{fig:stabVarKappa}a). Models with constant thermal conductivity predict only 6.6\% of the possible parameter combinations for Europa lead to a convecting ice shell. Including maximum feasible tidal heating with constant thermal conductivity increases the percentage to 28.5\%. For the preferred thermal conductivity relation presented here the percentage convecting without and with tidal heating is 1\% and 2.2\%, respectively. An increase in the thermal conductivity of ice near the surface lowers the probability of a convecting ice shell on Europa significantly, more so than including tidal heating increases the probability. Including tidal heating and constant thermal conductivity, as many previous studies have done \cite[for example][]{Mitri2005,Weller2019,Vilella2020}, substantially over predicts the likelihood of convection on Europa. In contrast, the choice of thermal conductivity relation minimally affects the predictions for convection in the ice shells of Enceladus and Titan. For example, in Enceladus' ice shell possible parameter combinations with no tidal heating that lead to convection decreases from 1.6\% to 0 when the constant conductivity model is replaced by the upper-bound \citet{Rabin2000} model. The opposite is the case for Titan, where most parameter combinations are convective irrespective of the thermal conductivity model chosen. 

The temperature dependence of thermal conductivity of ice Ih is not the only factor that can significantly affect the thermal conductivity of the ice shell. The near surface thermal conductivity of a tidally deforming ice shell is likely reduced by the presence of a porous and fractured regolith layer of several hundred meters to a few kilometers thick \citep{Nimmo2003}. Low thermal conductivity near the surface increases temperatures throughout the ice shell thereby favoring convection. Another important factor that could reduce the thermal conductivity of an ice shell is the formation of clathrates \citep{Ross1981}. Their effect on the thermal state of the ice shell depends on the location of formation. Near the surface the effect is similar to the porous regolith and increases the ice shell temperature \citep{Kalousova2020}. On the other hand, the delivery of gases from the ocean below could lead to the formation of hydrates near the base of the ice shell \citep{Kamata2019}. This would insulate the base of the ice shell, reduce the ice shell temperature and disfavor convection. Finally, the formation of hydrated salts can lower the conductivity of the ice shell \citep{Durham2010}. The effect on ice shell dynamics depends on both the concentration and location of the hydrated salts. On Europa hydrated salts have been observed at the surface \citep{Brown2013}.
The temperature-dependence of the thermal conductivity of ice Ih is therefore only one of several factors that must be taken into account. Interestingly, temperature appears to be the only factor that increases the thermal conductivity of the ice shell. 

The specific heat capacity of ice Ih at the surface is approximately one quarter of its value at the melting point (Figure~\ref{fig:iceProp}b). This reduces the energy stored in the conductive lid and will affect how the conductive lid responds to thermal perturbations (Figure~\ref{fig:cpRespTime}). Thermal perturbations can be induced by changes in tidal heating as an icy moon enters and exits resonances with other satellites \citep{Hussmann2004}, non-uniform delivery of internal heat due to ocean dynamics \citep{Soderlund2014}, and asteroid impacts to the surface \citep{Turtle2001}. The low heat capacity of the conductive lid may also ameliorate the outstanding problem of the formation of chaotic terrains on Europa. To our knowledge no process has been proposed that can deliver the large amounts of heat required to raise the temperature to the eutectic melting point and produce significant partial melting \citep{Collins2009}. For example, \citet{Nimmo2005} calculate the near surface melting induced by a warm sill or diapir using a specific heat of $\sim$2 kJ/kg/K, but Figure~\ref{fig:iceProp}b suggests near surface values are less than half of this value. This suggests that the same delivery of heat can raise a larger volume of ice to the eutectic melting point of the ice, though the reduction is likely not enough to induce substantial melting within the first 3 km beneath the surface, as inferred by \citet{Schmidt2011}.

Finally, the variation of the specific heat and thermal conductivity of ice Ih also strongly affect the performance of so-called cryobots that are currently being developed to penetrate the ice shell and sample the subsurface ocean \citep{Aamot1970}. Cryobots are cylindrical probes that penetrate the ice by melting and sink only due to gravity. Cryobot performance is affected in two ways. First, the thermal conductivity controls the geotherm in the ice shell and the temperature differential that must be overcome to induce melting. Second, specific heat and thermal conductivity directly control the descent rate of the probe for a given heating power \citep{Brandt2019}. The increased thermal conductivity of ice Ih at low temperature, proposed here, would decrease cryobot performance in the conductive lid due to larger lateral heat loss.

\section{Conclusion}
We use an enthalpy method to study ice shell convection with temperature-dependent material properties. Accounting for the temperature dependence of all material properties is important because the outer ice shells of icy ocean worlds experience an extremely large temperature range. We show that simulations assuming constant specific heat and thermal conductivity give potentially misleading results.

A review of the available thermal conductivity data for ice Ih shows that the commonly used Hobbs model represents a lower bound and likely underestimates the conductivity of ice. We present a new conductivity model for ice Ih that gives a best fit over the temperature range experienced in the ice shells of icy ocean worlds. Our model increases the thermal conductivity of ice at low temperatures by approximately 20\% compared to the Hobbs model. This results in up to an order of magnitude change in thermal conductivity across the ice shell. Whereas the viscosity of the ice shells of ocean worlds will likely remain an unknown until observations of the grain size at depth are obtained, the thermal conductivity of ice---particularly at the low temperatures relevant to ocean worlds---could be better constrained through experiments. At these temperatures the current variability in experimental data results in up to a factor of three change in the critical Ra number. Sample preparation factors such as freezing rate and aging should be addressed in future experiments to obtain more accurate measurements.

Our systematic study of ice shell stability shows that accounting for the increase in thermal conductivity near the surface has the potential to stabilize an ice shell. This can increase the critical Ra number for the onset of convection by an order of magnitude, which is larger than the impact of tidal heating on stability. The aforementioned is important because most published numerical simulations assume constant thermal conductivity and include tidal heating which amplifies the likelihood of convection on Europa more than ten fold. Thus, they overestimate the likelihood of convective mass and energy transfer. 

We show that the specific heat of ice Ih determines the amount of energy stored in the ice shell. The reduction of specific heat at low temperature reduces the energy stored in the conductive lid. Models accounting for temperature-dependent specific heat result in a faster response time to thermal perturbations, which may come in the form of changes in tidal heating or variations in ocean heat flow. This may help to explain surface features such as chaotic terrains which require large additions of energy to the near-surface ice.

\section{Acknowledgments}
E. C. was supported by a Jackson School Recruiting Fellowship and a Provost Office Supplemental Fellowship. N. S. W. was supported by the G. Unger Vetlesen Foundation and the Zonta International Amelia Earhart Fellowship. She also acknowledges the incredible support of the library staff who were able to track down ancient, long-lost texts containing measurements of the thermal conductivity of ice. J. S. J. would like to acknowledge support from the Earth, Environmental and Planetary Sciences Department at Rice University. M. A. H. was supported by National Aeronautics and Space Administration (NASA) Grant 18-EW18\_2-0027. M. A. H also acknowledges fruitful discussions with the students of GEO 325M in spring 2020 at UT Austin, who worked on part of this problem as a class project. 
All Matlab scripts that make up the numerical model presented here are given at \url{https://github.com/utImpacts/iceShellConvection} and are based on functions from the Matlab discrete Operator Toolbox, available at \url{https://github.com/mhesse/MatlabDiscreteOperatorToolbox/wiki}. All data used for the thermal conductivity fits and fit characteristics are available in the supplement. 
\bibliography{marc}
\end{document}


\begin{frontmatter}
\nolinenumbers
\title{Supporting Information for ``New insights into temperature-dependent ice properties and their effect on ice shell convection for icy ocean worlds"}

\author[jackson,utig,oden]{Evan Carnahan} 
\cortext[mycorrespondingauthor]{Corresponding author}
\ead{evan.carnahan@utexas.edu}

\author[jackson,utig]{Natalie S. Wolfenbarger}

\author[rice]{Jacob S. Jordan}

\author[jackson,oden]{Marc A. Hesse}

\address[jackson]{Department of Geological Studies, Jackson School of Geosciences, The University of Texas at Austin, Austin, TX, USA}
\address[utig]{Institute for Geophysics, Jackson School of Geosciences, The University of Texas at Austin, Austin, TX, USA}
\address[oden]{Oden Institute for Computational Science and Engineering, The University of Texas at Austin, Austin, TX, USA}
\address[rice]{Department of Earth, Environmental and Planetary Sciences, Rice University, Houston, Texas, USA}

\end{frontmatter}

%
%





%
%

\noindent\textbf{Contents of this file}
\nolinenumbers
\begin{enumerate}
\item Text: Review of thermal conductivity data of ice Ih, dimensional equations, dimensionless equations, numerical solution, and simulations
\item Figures 1 to 5
\item Tables 1 to 2
\end{enumerate}
\noindent\textbf{Additional Supporting Information (Files uploaded separately)}
\begin{enumerate}
\item Captions for Datasets 1
\end{enumerate}

\section{Review of the thermal conductivity data of ice Ih}
\doublespacing
The thermal conductivity of ice Ih as a function of temperature is relevant to a number of applications in a variety of fields. The model of \citet{Hobbs1974} is commonly used in planetary science whereas the model of \citet{Rabin2000} is favored in cryosurgery research. Although these models represent best fits to particular datasets, they do not represent the best fit to the full spectrum of available data. For example, the Hobbs model was obtained by fitting two datasets: the data of \citet{Jakob1929} and \citet{Dillard1966}. Later publications noted the data were likely to be in error \citep{Dillard1969,Slack1980}; however examination of the data in context of the more recent datasets does not suggest the data are out-of-family. The Rabin model includes the data of \citet{Jakob1929} but also incorporates the data of \citet{Ratcliffe1962} and \citet{Dean1963}. The data of \citet{Dean1963} were described as possibly 10\% too high by \citet{Hobbs1974}. 

\begin{figure}
    \centering
    \includegraphics[width = \textwidth]{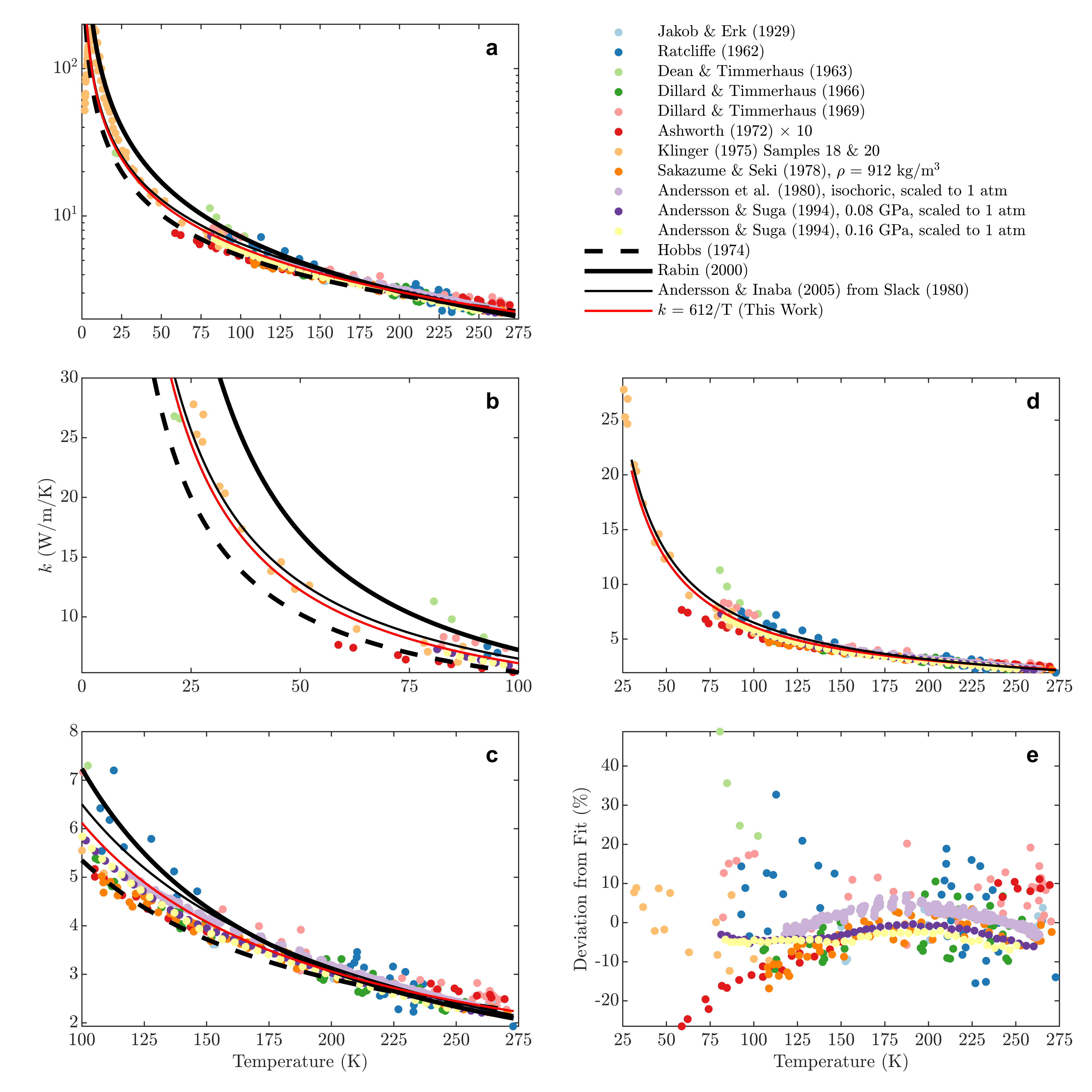}
    \caption{Summary of published thermal conductivity data and select published models, including our proposed fit: (a) spanning full temperature range of stable ice Ih (b) spanning the temperature range 0 $\leq$ T $\leq$ 100 K (c) spanning the temperature range 100 $\leq$ T $\leq$ 273 K (d) spanning the temperature range relevant to icy ocean worlds, 30 $\leq$ T $\leq$ 273 K. (e) Residuals, expressed as a percent deviation of the data relative to our fit, spanning the temperature range relevant to icy ocean worlds, 30 $\leq$ T $\leq$ 273 K.}
    \label{fig:k_summary}
\end{figure}

To obtain a model that best represents the thermal conductivity of ice, we fit a collection of the available published data (see Figure~\ref{fig:k_summary}d), excluding the isochoric data of \citet{Andersson1994} and instead including their isobaric data at 0.08 GPa and 0.16 GPa, corrected to atmospheric pressure using $\frac{\partial{\ln{k}}}{\partial{P}}=-0.28$  GPa$^{-1}$. The data of \citet{Andersson1980} is corrected from isochoric to isobaric conditions, by accounting for the change in density expected with temperature using $\frac{\partial{\ln{k}}}{\partial{\ln{\rho}}}=-2.6$ from \citet{Andersson1980} and the volumetric thermal expansion coefficient for ice as a function of temperature obtained by \citet{Leadbetter1965} and presented in \citet{Hobbs1974}. The previous comprehensive analysis to obtain the "best estimate" of the thermal conductivity of ice was performed by \citet{Slack1980}. His "best estimate" datapoints were used by \citet{Andersson2005} to obtain a model of the form $k = a/T+b+cT$. Notably, their coefficients $b$ and $c$ are small relative to $a$ and contribute minimally to the magnitude of thermal conductivity, suggesting a model of the form $k = a/T$ is sufficient to represent the thermal conductivity of ice over the range of temperatures considered here. As such, we adopt a model where thermal conductivity is inversely proportional to temperature. 

We find that the model $k = 612/T$ fits the data well over temperatures ranging from 30 to 273 K (Data Set S1). Selectively excluding particular datasets generally resulted in changes to the best fit coefficient within $\pm5$. Excluding the data of \citet{Ashworth1972} had the largest effect (resulting in a best fit coefficient of $619$) because of the low values of thermal conductivity measured at low temperatures. The deviation from the model, expressed in a percentage of the model value, is provided in Figure \ref{fig:k_summary}. We recognize our model is heavily biased to the data of \citet{Andersson1980}, which is comprised of over 400 datapoints whereas the other datasets include less than 50 datapoints. The "best estimate" datapoints of \citet{Slack1980} appear similarly biased towards the dataset of \citet{Andersson1980}. The deviation for most datasets is within 10\%; however, certain datasets exceed the model prediction by over 20\% at temperatures below approximately 100 K. In particular, the data of \citet{Dean1963} appears out-of-family relative to other datasets and consistently exceeds the model prediction by 20\%, supporting the interpretation of \citet{Hobbs1974} that these measurements may be too high. The estimates for our model accuracy are consistent with those claimed by \citet{Slack1980}. We note that we include a number of datasets not considered by \citet{Slack1980} including the high temperature data ($>$200 K) presented in \citet{Choi1985}, the data of \citet{Ashworth1972} which span in temperature from approximately 140 to 60 K (corrected by a factor of 10), the data of \citet{Sakazume1978}, and the more recent data of \citet{Andersson1994} between 260 and 80~K. Examination of the available published data shown in Figure \ref{fig:k_summary} reveals that although certain datasets appear to have high precision, particularly the datasets of \citet{Ashworth1972}, \citet{Sakazume1978}, \citet{Andersson1980}, and \citet{Andersson1994}, the accuracy is less well-constrained. These datasets appear to deviate significantly from each other at temperatures lower than 100 K. At the melting temperature, our model predicts a thermal conductivity of 2.24 W/m/K which is in-family with the value predicted by other models ($\sim$2-3 W/m/K). Fit characteristics for the models of thermal conductivity used in this study, including R$^2$ values and ranges of validity, evaluated from all data referenced in this work are given in Data Set S1. 

Both recent and past studies suggest that the thermal conductivity of ice is highly sensitive to the sample preparation method. \citet{Klinger1975} found that ice grown in Teflon vessels presented with higher thermal conductivity than ice grown in plexiglass vessels for temperatures below 20 K. Higher growth rates resulted in lower thermal conductivity for the same temperature range. \citet{Bonales2017} examined discrepancies in measured thermal conductivity at temperatures greater than approximately 230 K and found higher freezing rates resulted in higher thermal conductivity by up to 10\%. Their data for slowly frozen samples is consistent with the data in \citet{Choi1985}, the values of which are within 2\% of our proposed model. The correlation to freezing rate observed by \citet{Bonales2017} is notably opposite to the conclusion of \citet{Klinger1975} for much lower temperatures. Still, these studies suggest differences in freezing rates between experiments could be responsible for introducing the observed discrepancies between datasets. The influence of anisotropy on thermal conductivity is thought to be capable of introducing variations of approximately 5\% \citet{Klinger1975}. Later work found the discrepancy could reach 20\% but only at temperatures between 2 and 10 K \citep{Klinger1982}. Aging of the ice was also found to lower the thermal conductivity by up to 30\% for temperatures below 10 K, due to defect migration \citep{Klinger1982}. 

\section{Dimensional model equations}
The model for ice shell convection is comprised of a set of conservation equations and constitutive relations. The values of all model parameters and the units are given in Table~\ref{tab:SymChar}.
\subsection{Constitutive relations}
Here we give the expressions we have used for the temperature-dependence of the material properties of ice Ih. We explore the following relationships for $k (T)$
\begin{subequations}
\begin{align}
    k (T) & = 2.26\\
    k (T) &= 0.4685 + 488.12/T\\
  k(T) &= 2135/T^{1.235}\\
  k(T) &= 612/T
\end{align}
\end{subequations}
all given in $\left[ \frac{\text{W}}{\text{m K}}\right]$, \cite{Hobbs1974,Rabin2000}. For the specific heat we use the expression of \citet{Ellsworth1983}, 
\begin{linenomath*}
\begin{equation} \label{eqn:supC_p}
    c_p(T) = 185+7.037\,T\quad\mbox{in}\quad \left[ \frac{\text{J}}{\text{kg K}} \right].
\end{equation}
\end{linenomath*}
We assume a Newtonian fluid and  an Arrhenius dependence of the dynamic viscosity on temperature given by 
\begin{linenomath*}
\begin{align}
    \eta = \eta_c \ \text{exp}\left[A \left(\frac{T_m}{T} - 1\right)\right]\quad\mbox{in}\quad\left[\text{Pa s}\right],
\end{align}
\end{linenomath*}
where $A = \frac{E_a}{R T_m}$, $E_a$ is the activation energy, $R$ is the universal gas constant, $\eta_c$ is the viscosity at the melting temperature, and $T_m$ is the melting temperature of ice \citep{Goldsby2001}. Finally, we use the density of ice derived from the Gibbs potential function of \citet{Feistel2006}, which can be approximated within the temperature range of our study by the expression
\begin{linenomath*}
\begin{align}
    \rho = -3.621\times10^{-4}T^{2}+3.977\times10^{-2}T+932.7\quad\mbox{in}\quad\left[\text{kg/m}^3\right].
\end{align}
\end{linenomath*}
These material properties for ice Ih, except viscosity, are plotted in Figure 1 of the main manuscript together with select data sets. 
\subsection{Conservation laws}
Conservation of momentum for a highly viscous fluid is given by the Stokes Equation
\begin{linenomath*}
\begin{equation} \label{eqn:supNS}
    \nabla \cdot [\eta (\nabla \mathbf{u}+\nabla \mathbf{u}^T)]-\nabla P = (\rho - \rho_c) g \hat{\mathbf{z}}
\end{equation}
\end{linenomath*}
where $P=p+\rho_cgz$ ($\nabla P = \nabla p + \rho_c g \hat{\mathbf{z}}$) is the modified pressure, $\mathbf{u}$ is the velocity vector, $\rho$ is the density, $\eta$ is the dynamic viscosity, $g$ is gravity, $P$ is pressure without the hydrostatic component at the characteristic density $\rho_c$, and $\eta$ is the viscosity. 
We apply the Oberbeck-Boussinesq approximation and only treat density variations in the body force of \eqref{eqn:supNS}, \citep{Oberbeck1879,Boussinesq1903}. In this limit, the conservation equation for fluid mass is divergence free 
\begin{linenomath*}
\begin{equation} \label{eqn:supCOM}
    \nabla \cdot \mathbf{u} = 0
\end{equation}
\end{linenomath*}
Conservation of energy with no change in volume, Oberbeck-Boussinesq approximation, is equivalent to conservation of volumetric enthalpy, $H$. Volumetric enthalpy,
\begin{linenomath*}
\begin{align}\label{eqn:supEnthalpy}
    H=\rho \left[ h_0 + \int_{T_0}^T c_p(\tau) d\tau \right]
\end{align}
\end{linenomath*}
where $h_0$ is the reference enthalpy associated with temperature $T_0$, and $c_p$ is the specific heat. Choosing $T_0=T_m$ and $h_0=0$ and the linear expression for $c_p = a+b T$, the enthalpy of the ice is given by
\begin{linenomath*}
\begin{align} \label{eqn:TtoH}
    H = \rho_0 \left(a (T-T_m)+\frac{b}{2} (T^2-T_m^2)\right),
\end{align}
\end{linenomath*}
where $\rho=\rho_0$ due to the Oberbeck-Boussinesq approximation. Conservation of volumetric enthalpy is given by
\begin{linenomath*}
\begin{equation} \label{eqn:supCOE}
    \frac{\partial H}{\partial t} +\nabla \cdot\left[\mathbf{u} H - k(T) \nabla T \right] = G
\end{equation}
\end{linenomath*}
where $T$ is temperature, $G$ is the energy source, and $k$ is the thermal conductivity. This equation is in terms of $H$ and $T$, but it can be written in terms of $H$ only as
\begin{linenomath*}
\begin{equation} \label{eqn:supCOE_H}
    \frac{\partial H}{\partial t} +\nabla \cdot\left[\mathbf{u} H - \kappa(H) \nabla H \right] = G,
\end{equation}
\end{linenomath*}
where the thermal diffusivity is given by
\begin{linenomath*}
\begin{align}
    \kappa(H) = k(T(H))\frac{dT}{dH}.
\end{align}
\end{linenomath*}
Here $k(T)$ is evaluated in terms of $H$ by substituting
\begin{linenomath*}
\begin{align}\label{eqn:T(H)}
    T = \frac{-a + \sqrt{a^2 + 2b (aT_m + H/\rho_0 + \frac{b}{2} T_m^2)}}{b}
\end{align}
\end{linenomath*}
and 
\begin{linenomath*}
\begin{align}
    \frac{\mathrm{d}T}{\mathrm{d}H} = \frac{1}{\rho c_{p,c} \sqrt{1+\frac{2b}{\rho (c_{p,c})^2 H}}}.\label{eqn:dTdH}
\end{align}
\end{linenomath*}
The coefficients $a$ and $b$ in equations \eqref{eqn:T(H)} and \eqref{eqn:dTdH} are from the $c_p$ property relation (Table~\ref{tab:prop_coeff}).

We model the energy source term from tidal dissipation. Following \citep{Tobie2003}, 
\begin{linenomath*}
\begin{align}
    G = \frac{2 \eta G_{max}}{1 + \eta^2}
\end{align}
\end{linenomath*}
where $G_{max}$ is the maximum tidal heating rate. As we model a general ice shell we assume that the maximum tidal heating occurs at the melt viscosity \citep{Tobie2003,Vilella2020}.
\section{Dimensionless model equations}
We introduce the characteristic scales for the temperature-dependent material properties by evaluating them at the basal temperature
\begin{linenomath*}
\begin{align}
    k_c = k(T_b),\ \  c_{p,c} = c_p(T_b),\ \  \eta_c = \eta(T_b),\ \  \rho_c = \rho(T_b),\ \  G_c = G_{max},  \ \  \mathrm{and}\ \ \kappa_c= \frac{k_c}{\rho_c c_{p,c}}.
\end{align}
\end{linenomath*}
Numeric values of these characteristic quantities are given in Table~\ref{tab:SymChar}. Based on these characteristic property values we can define the following characteristic scales for the model variables
\begin{linenomath*}
\begin{align}
    x_c = d,\quad t_c = \frac{x_c^2}{\kappa_c},\quad u_c = \frac{x_c}{t_c},\quad P_c = \frac{\eta_c \kappa_c}{x_c^2}, \quad \text{and} \quad H_c = \rho_c c_{p,c}\Delta T
\end{align}
\end{linenomath*}
where $\Delta T = T_b-T_s$ and $d$ is the ice shell thickness. Given these scales we introduce the dimensionless variables
\begin{linenomath*}
\begin{align}
    x'=\frac{\mathbf{x}}{x_c},\  t' = \frac{t}{t_c},\  \mathbf{u}' = \frac{\mathbf{u}}{u_c},\  P' = \frac{P}{P_c},\  H' = \frac{H}{H_c},\  T' = \frac{T-T_s}{\Delta T},  \ \rho' = \frac{\rho - \rho_b}{\rho_b - \rho_t}, \ G' = \frac{G}{G_{max}}.
\end{align}
\end{linenomath*}
Below we first present the dimensionless constitutive relations for material properties followed by the dimensionless conservation laws for linear momentum, mass, and energy.

\subsection{Dimensionless constitutive relations}

The parameterizations of the temperature-dependence of the specific heat, the thermal conductivity and the exponent of the Arrhenius relation for viscosity can be written as a power-law plus a constant,
\begin{linenomath*}
\begin{align} \label{eqn:supPhysFunc}
    f = a+b T^v.
\end{align}
\end{linenomath*}
We thus non-dimensionalize this general form to show that it is only dependent on the homologous surface temperature, $\Theta=T_s/T_m$, which controls the contrast of the dimensionless material property, $f'$, across the ice shell. The dimensionless form is
\begin{linenomath*}
\begin{align}
        f'(T',\Theta) &= \frac{f(T')}{f(1)}= \pi_1 + \pi_2 ((1-\Theta)T'+\Theta)^{\pi_3},
\end{align}
where the coefficients 
\begin{align}
\pi_1=\frac{a}{a+bT_m^v}, \quad \pi_2=\frac{bT_m^v}{a+bT_m^v}, \quad \text{and} \quad \pi_3=v 
\end{align}
\end{linenomath*}
are constant as long as the melting temperature, $T_m$, is constant. The only dimensionless group introduced by the temperature dependence is therefore the homologous temperature at the surface, $\Theta$. Both dimensional and dimensionless coefficients for the temperature-dependent material properties used here are given in Table~\ref{tab:prop_coeff}.

Comparing the non-dimensional groups for thermal conductivity of Rabin to our model, $\pi_1 = 0$, for both but $v$ is larger for Rabin. Comparing our thermal conductivity to Hobbs, $\pi_3 = -1$ for both, however $\pi_2$ for Hobbs is less than one and for our model is 1. This demonstrates the order from most to least sensitive thermal conductivity relationship to changes in the homologous temperature, Rabin, our model, and Hobbs.

\subsection{Dimensionless conservation laws}
Using the scales introduced above the dimensionless equations are given by
\begin{linenomath*}
\begin{subequations}\label{eqn:gov - nondim}
\begin{align}
     \nabla' \cdot [ \eta' (\nabla' \mathbf{u}' + \nabla' \mathbf{u}'^T)]-
    \nabla' P' &= - \textrm{Ra}\, \rho'\hat{\mathbf{z}},\label{eqn:mom non-dim}\\
    \nabla \cdot \mathbf{u}' &= 0,\label{eqn:mass non-dim}\\
     \frac{\partial H'}{\partial t'} + \nabla' \cdot \left[\mathbf{u}'H' - \kappa'(H')  \nabla' H'\right] &=  \tidal\, \frac{2 \eta'}{1 + (\eta')^2}.\label{eqn:energy non-dim}
\end{align}
\end{subequations}
\end{linenomath*}

The dimensionless thermal diffusivity is given by
\begin{linenomath*}
\begin{align}
    \kappa'(H') = k'(T'(H')) \frac{\mathrm{d} T'}{\mathrm{d} H'} = \frac{k'(T'(H'))}{\sqrt{1 + 2 \pi_2 (1 - \Theta)\, H'}},
\end{align}
\end{linenomath*}
where $\pi_2$ is evaluated with constants from the $c_p$ property relation (Table~\ref{tab:prop_coeff}) and $k'(T')$ is evaluated as function of $H'$ using the relation
\begin{linenomath*}
\begin{align}
   T'(H') =  \frac{-a - b T_s + \sqrt{c_{p,c}} \sqrt{2b\Delta T H' + c_{p,c}}}{b \Delta T} .
\end{align}
\end{linenomath*}

The dimensionless model equations are dependent on the following dimensionless parameters
\begin{linenomath*}
\begin{align}
    \label{eqn:nonParam}
    \textrm{Ra} = \frac{g  \Delta \rho\, d^3}{\eta_c \kappa_{c}}, \quad 
    \tidal = \frac{G_{\text{max}}\, d^2}{k_c \Delta T}, 
    \quad \Theta = \frac{T_s}{T_b}, \quad
    \mathrm{and}
    \quad \mathrm{Ar}=\frac{w}{d}.
\end{align}
\end{linenomath*}
where $w$ is the width of the domain.

\section{Numerical solution}
The numerical solution of the governing equation requires a spatial discretization, a temporal discretization, and a strategy to handle the non-linearity of the governing equations. Each aspect is described separately below.

\subsection{Staggered grids}
\begin{figure}
    \centering
    \includegraphics[width = \textwidth]{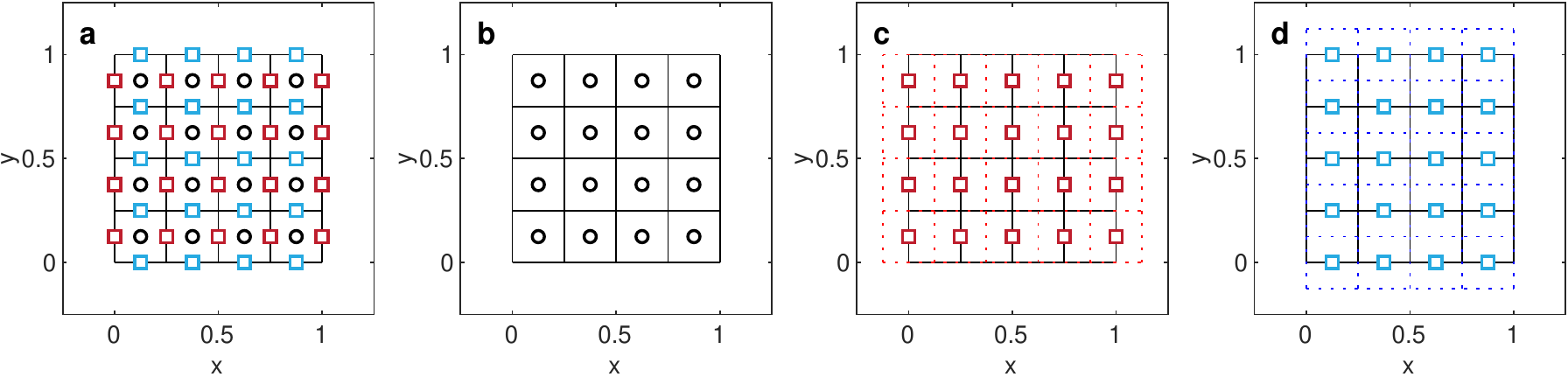}
    \caption{a) Staggered grid for the discretization of the Stokes equation showing the location of all variables. black: pressure and enthalpy. red: $x$-velocity component. blue: $y$-velocity component . b) The pressure grid is the primary grid. Pressure unknowns shown as black circles. c) Grid for the horizontal component of the velocity, $v_x$, is shown as a red dashed line. It is shifted by $\Delta x/2$ in $x$-direction relative to the primary grid. The $x$-velocity unknown are shown as red squares.  d) Grid for the vertical component of the velocity, $v_y$, is shown as a blue dashed line. It is shifted by $\Delta y/2$ in $y$-direction relative to the primary grid. The $y$-velocity unknown are shown as blue squares.}
    \label{fig:grids}
\end{figure}

We employ a conservative finite difference discretization to ensure discrete conservation of mass, momentum, and energy \citep{LeVeque1992}. This requires a staggered mesh, shown in Figure~\ref{fig:grids}a, with pressure and enthalpy in the cell center and velocity components on the corresponding cell faces \citep{Harlow1965}. Due to the square domain, $\left[0,\,\mathrm{Ar}\right]\times\left[0,\,1\right]$ we use a simple Cartesian tensor product mesh with twice as fine resolution in the $y$-direction. For $\mathit{Nx}$ cells in the $x$-direction and $\Ny$ cells in the $y$-direction the total number cells is $N=\Nx\cdot \Ny$ and cell size is given by $\Delta x = \mathrm{Ar}/\Nx$ and $\Delta y = 1/\Ny$. The number of $x$-faces is $\Nfx = (\Nx+1)\cdot Ny$ and the number of $y$-faces is $\Nfy = \Nx\cdot (\Ny+1)$.

To simplify the implementation we define three different  staggered grids. The primary grid is for scalar unknowns such as pressure and enthalpy (Figure~\ref{fig:grids}b) and will be referred to as the pressure gird below. The grids for the velocity components, $u_x$ and $u_y$, are centered around their respective unknown and hence shifted by $\Delta x/2$ and $\Delta y/2$ (Figure~\ref{fig:grids}c and~\ref{fig:grids}d). The cell size, $\Delta x$ and $\Delta y$, in all three grids is the same but the grid sizes change as follows
\begin{linenomath*}
\begin{subequations}
\begin{align}
    \mbox{pressure grid:}     && \Nx &          && \Ny           && N = \Nx\cdot \Ny\\
    \mbox{$x$-velocity grid:} && \Nx_x &= \Nx+1 && \Ny_x = \Ny   && N_x = (\Nx+1)\cdot \Ny\\
    \mbox{$y$-velocity grid:} && \Nx_y &= \Nx   && \Ny_y = \Ny+1 && N_y = \Nx\cdot (\Ny+1)
\end{align}
\end{subequations}
\end{linenomath*}
\subsubsection{Operator implementation of the conservative finite difference discretization}\label{sec:tensor product}

The implementation is facilitated by directly discretizing the basic differential operators of vector calculus: divergence, gradient, and curl. This is similar to mimetic finite difference methods \citep{daVeiga2014,Haber2015}, though without the complications introduced by irregular grids. For the system of equations \eqref{eqn:gov - nondim} only the discrete divergence, $\D$, and the discrete gradient, $\G$, are needed. For a scalar unknown on a regular Cartesian mesh, as in Figure~\ref{fig:grids}b, the two-dimensional discrete operators used here have been given by \cite{Hesse2019}. We extend this approach to the discretization of the gradient of a vector and the divergence of a tensor which are required for the discretization of the Stokes equation with variable viscosity on the staggered grid shown in Figure~\ref{fig:grids}.

\subsubsection{Discrete divergence and gradient for scalar unknown}\label{eqn:D and G}
Following supplementary materials of \cite{Hesse2019} the two-dimensional discrete divergence and gradient on the staggered mesh in Figure~\ref{fig:grids}a are constructed using the tensor product
\begin{linenomath*}
\begin{align}
    \D = \begin{bmatrix}\D_x\otimes\I_y& \I_x \otimes \D_y\end{bmatrix}\quad\mbox{and}\quad\G =-\D^T\label{eqn:D and G}.
\end{align}
\end{linenomath*}
Here the two dimensional divergence, $\D$, is generated from the one dimensional operators, $\D_x$ and $\D_y$, using tensor products and the discrete two-dimensional gradient, $\G$, is obtained from the discrete adjoint relationship. The matrices $\I_x$ and $\I_y$ are square identity operators of size $\Nx$ and $\Ny$, respectively. On the boundaries the gradient is set to zero, so that natural boundary conditions are imposed unless specified otherwise. \cite{Hesse2019} give expressions for the central difference discretization of the one dimensional operators $\D_x$ and $\D_y$ in their supplementary materials. This construction can be extended to three-dimensional discretizations through the application of a second tensor product \citep{Haber2015}.

For the Stokes discretization we use three staggered grids and each comes with a corresponding set of discrete operators that will be denoted as follows:
\begin{linenomath*}
\begin{subequations}
\begin{align}
    \mbox{pressure grid:} && \Dp=\begin{bmatrix}\Dpx & \Dpy\end{bmatrix}\quad\mbox{and}\quad \Gp=\begin{bmatrix}\Gpx \\ \Gpy\end{bmatrix}\\
    x\mbox{-velocity grid:} && \Dx=\begin{bmatrix}\Dxx & \Dxy\end{bmatrix}\quad\mbox{and}\quad \Gx=\begin{bmatrix}\Gxx \\ \Gxy\end{bmatrix}\\
    y\mbox{-velocity grid:} && \Dy=\begin{bmatrix}\Dyx & \Dyy\end{bmatrix}\quad\mbox{and}\quad \Gy=\begin{bmatrix}\Gyx \\ \Gyy\end{bmatrix}
\end{align}
\end{subequations}
\end{linenomath*}
Here we have identified the matrix blocks that apply the operator in the $x$ and $y$ directions respectively. For example, $\Dxy$ computes the divergence in the $y$-direction of the $x$-velocity grid and $\Gpx$ computes the gradient in the $x$-direction on the pressure grid.

\subsection{Stokes system matrix}
To discretize the Stokes equation (\ref{eqn:mom non-dim} and \ref{eqn:mass non-dim}) we order the unknown vector as, $\begin{bmatrix}
\mathbf{u} & p 
\end{bmatrix}^T$, so that the discrete Stokes system takes the form
\begin{linenomath*}
\begin{align}
    \begin{bmatrix}
\SS & -\Gp \\
\Dp & \mathbf{0}
\end{bmatrix}\begin{bmatrix}
\mathbf{u}\\
\mathbf{p}
\end{bmatrix}=\begin{bmatrix}
\mathbf{f}\\
\mathbf{0}
\end{bmatrix}
\end{align}
\end{linenomath*}
where the first row corresponds to \eqref{eqn:mom non-dim} and the second row to \eqref{eqn:mass non-dim}. 
To discretize the Stokes system we therefore need to find the expression for the  matrix $\SS$, which discretizes the divergence of the deviatoric stress tensor, $\SS \approx \nabla \cdot \dev$, where we have dropped the primes. The deviatoric stress tensor, $\dev = 2 \eta \epsdot$, is a linear function of the strain rate (rate of deformation) tensor
\begin{linenomath*}
\begin{align}
    \epsdot = \frac{1}{2}\left( \nabla \u + \nabla \u^T\right),
\end{align}
\end{linenomath*}
where we have also dropped the primes. Computing $\SS$ therefore requires three steps
\begin{enumerate}
    \item Compute the discrete strain rate tensor: $\epsdot \approx \edot$ (stored as vector).
    \item Compute the discrete deviatoric stress tensor: $\dev = \devvec$ (stored as vector).
    \item Compute the divergence of the deviatoric stress: $\nabla\cdot\dev = \D*\devvec$.
\end{enumerate}

\subsubsection{Discrete strain rate tensor}
The discrete strain rate tensor is a linear function of the velocity so that it can be computed as the following matrix vector product
\begin{linenomath*}
\begin{align}
    \edot = \Edot\cdot\u,
\end{align}
\end{linenomath*}
where $\u = \begin{bmatrix}\u_x & \u_y \end{bmatrix}^T$ is the vector of all discrete velocities ordered by direction. The components of the strain rate tensor are 
\begin{linenomath*}
\begin{align}
    \epsdot = \begin{bmatrix}
u_{x,x} & \frac{1}{2}\left(u_{1,2}+u_{2,1}\right) \\
\frac{1}{2}\left(u_{1,2}+u_{2,1}\right) & u_{y,y}
\end{bmatrix}=\begin{bmatrix}
\dot\epsilon_{xx} & \dot\epsilon_c \\
\dot\epsilon_c & \dot\epsilon_{yy}
\end{bmatrix}
\end{align}
\end{linenomath*}
and we store the discrete approximations of the three independent components in the vector 
\begin{linenomath*}
\begin{align}
    \edot = \begin{bmatrix}
\edot_{xx} \\ \edot_{yy} \\ \edot_c\end{bmatrix}.
\end{align}
\end{linenomath*}
Here $\edot_{xx}$ and $\edot_{yy}$ are vectors of all discrete appoximations of $\dot\epsilon_{xx}$ and $\dot\epsilon_{yy}$, computed at the cell centers of the pressure grid. Similarly, $\edot_c$ is a vector of all discrete approximations of $\dot\epsilon_c$, but computed at the cell corners of the pressure grid. Given the ordering of $\edot$ and $\u$ the matrix that computes the strain rates from the velocities is
\begin{linenomath*}
\begin{align}
\Edot = 
\begin{bmatrix}
    \Gxx & \Zxy  \\
    \Zyx & \Gyy \\
   \frac{1}{2}\Gxy &\frac{1}{2}\Gyx&
\end{bmatrix},
\end{align}
\end{linenomath*}
where $\Zxy$ and $\Zyx$ are zero matrices of appropriate size. 

\subsubsection{Discrete deviatoric stress tensor}
The discrete approximation of the deviatoric stress tensor
\begin{linenomath*}
\begin{align}
    \dev = \begin{bmatrix}
        \tau_{xx} & \tau_{c}\\ \tau_c & \tau_{yy}
    \end{bmatrix}
\end{align}
\end{linenomath*}
is stored as a vector similar to the strain rate tensor
\begin{linenomath*}
\begin{align}
    \devvec = \begin{bmatrix}
        \devvec_{xx}\\
        \devvec_{yy}\\
        \devvec_c
    \end{bmatrix},
\end{align}
\end{linenomath*}
where $\devvec_{xx}$ and $\devvec_{yy}$ are vectors of all discrete approximations of $\tau_{xx}$ and $\tau_{yy}$, computed at the cell centers of the pressure grid. Similarly, $\devvec_c$ is a vector of all discrete approximations of $\tau_c$, but computed at the cell corners of the pressure grid.
For an isoviscous fluid the discrete approximation of the deviatoric stress is simply
\begin{linenomath*}
\begin{align}
    \devvec = 2\eta\, \edot.
\end{align}
\end{linenomath*}
If the viscosity depends on temperature, $\eta = \eta(T)$, it is spatially variable. The temperature is computed in the centers of the pressure cells and hence needs to be averaged to the faces and corners of the pressure cells. In this case, the deviatoric stress is given by
\begin{linenomath*}
\begin{align}
    \devvec = 2\,\Etad\,\edot,
\end{align}
\end{linenomath*}
where $\Etad$ is a matrix that contains the viscosities averaged to the appropriate locations. The temperature at cell centers are arithmetically averaged to the cell faces and corners then the temperature dependent viscosity is calculated at those locations. 

\subsubsection{Discrete divergence of the deviatoric stress}
Finally, we need to compute the divergence of the deviatoric stress tensor. The divergence is applied to each row of the tensor so that
\begin{linenomath*}
\begin{align}
    \nabla\cdot\dev = \begin{bmatrix}\nabla\cdot\begin{bmatrix}
        \tau_{xx} & \tau_{c}
    \end{bmatrix}\\
       \nabla\cdot\begin{bmatrix}
        \tau_{c} & \tau_{yy}
    \end{bmatrix}
    \end{bmatrix} = \begin{bmatrix}
        \tau_{xx,x} & \tau_{c,y}\\
        \tau_{c,x} & \tau_{yy,y}
    \end{bmatrix}.
\end{align}
\end{linenomath*}
The discrete approximation is given by
\begin{linenomath*}
\begin{align}
    \nabla\cdot\dev \approx \D\, \devvec,
\end{align}
\end{linenomath*}
where the tensor divergence is given by
\begin{linenomath*}
\begin{align}
    \D = \begin{bmatrix}
        \Dxx & \Zyx^T & \Dxy \\
        \Zxy^T & \Dyy & \Dyx
    \end{bmatrix},
\end{align}
\end{linenomath*}
in terms of the discrete operators on the grids for the $x$ and $y$ velocities.

The matrix $\SS$ in the Stokes system is therefore given by
\begin{linenomath*}
\begin{align}
    \SS = 2\, \D\,\Etad\,\Edot.
\end{align}
In the isoviscous case 
\begin{align}
    \A = 2\eta\,\D\,\Edot = 2\eta\, \begin{bmatrix}
	\Dxx\cdot\Gxx +  \frac{1}{2}\Dxy\cdot\Gxy &
	\frac{1}{2} \Dxy\cdot\Gyx\\
	\frac{1}{2}\Dyx\cdot\Gxy &
	\Dyy\cdot\Gyy+ \frac{1}{2}\Dyx\cdot\Gyx \\
\end{bmatrix},
\end{align}
\end{linenomath*}
so that $\SS=\SS^T$ and the Stokes system is symmetric, except for terms modified by natural boundary conditions. 

\subsection{Discretization of the advection-diffusion equation}
\subsubsection{Time-stepping}
We discretize the energy equation \eqref{eqn:energy non-dim} using a second-order Crank-Nicholson time-stepping scheme
\begin{linenomath*}
\begin{align} \label{eqn:CN}
        \left(\Ip + \frac{\Delta t}{2} \L\right) \H^{n+1} = \left(\Ip - \frac{\Delta t}{2} \L\right) \H^{n} + \Delta t \tidal \frac{2 \etavec^2}{1+\etavec^2}
\end{align}
\end{linenomath*}
where $\Ip$ is the identity on the pressure grid, $\H$ is the vector of unknown enthalpies, $\etavec$ is the vector of viscosities at the pressure grid centers, and $\L$ is the steady advection-diffusion operator given by
\begin{linenomath*}
\begin{align}
    \L = \Dp\,\left(\A(\u) - \D_d(\H) \Gp\right).
\end{align}
\end{linenomath*}
Here $\Dp$ and $\Gp$ are the discrete divergence and gradient on the pressure grid, $\D_d$ is a diagonal matrix containing the averages of the thermal conductivity on the cell faces and $\A(\u)$ is the advection matrix. The velocities, $\u$, are given by the solution of the Stokes equation and the dependence of the thermal diffusivity and viscosity on enthalpy, $\H$, are lagged to the previous time step.

\subsubsection{Advection matrix}
The advection matrix, $\A(\u)$, is based on the upwind flux as determined by the ice velocity \citep{LeVeque1992}. Unlike the discrete vector calculus operators, $\D$ and $\G$ in \eqref{eqn:D and G}, the entries in $\A(\u)$ are not fixed and change as $\u$ changes. Hence $\A(\u)$ must be recomputed at every time step. As such we separate the potential locations of the entries, which are fixed from the associated magnitudes
\begin{linenomath*}
\begin{align}\label{eqn:adv mat}
    \A(\u) = \Ud^+(\u)\,\A^+ + \Ud^-(\u)\, \A^-, 
\end{align}
\end{linenomath*}
where $\A^+$ and $\A^-$ contain the locations of the positive and negative velocities respectively and $\Ud^+$ and $\Ud^-$ are diagonal matrices containing positive and negative velocity values, respectively. 

For a one dimensional grid with $\Nx$ cells and $\Nfx=\Nx+1$ faces, the location matrices are size $\Nfx\times\Nx$ and given by
\begin{linenomath*}
\begin{align}
    \A^+ = \begin{bmatrix}
    0    &        &         &  \\
    1    &       &      &\\
           & \ddots &  &    \\
           &        & 1 &   \\
           &        &         & 1
\end{bmatrix}\quad\mbox{and}\quad
\A^- = \begin{bmatrix}
    1 &           &         &  \\
       & 1        &      &\\
       &      &  \ddots &    \\
       &            &  &  1 \\
       &            &         & 0
\end{bmatrix}.
\end{align}
\end{linenomath*}
The diagonals of the $\Nfx\times\Nfx$ matrices $\Ud^+$ and $\Ud^-$ are given by $\max(\mathbf{u},0)$ and $\min(\mathbf{u},0)$, respectively. 

In two dimensions the overall advection matrix is assembled as in \eqref{eqn:adv mat}, but in the location matrices now contain two blocks
\begin{linenomath*}
\begin{align}
    \A^+ = \begin{bmatrix}
        \Ax^+\\\Ay^+
    \end{bmatrix}\quad \mbox{and}\quad \begin{bmatrix}
        \Ax^-\\\Ay^-
    \end{bmatrix}
\end{align}
\end{linenomath*}
for the velocities in the $x$ and $y$ directions. Similar to Section~\ref{sec:tensor product} these two-dimensional location matrices can be assembled from the one-dimensional location matrices using tensor products.
\begin{linenomath*}
\begin{align}
    \Ax^+ = \A_x^+\otimes \I_y \quad \mbox{and} \quad \Ax^- = \A_x^-\otimes \I_y,\\
    \Ay^+ = \I_x\otimes\A_y^+ \quad \mbox{and} \quad \Ay^- = \I_x \otimes \A_y^-.
\end{align}
\end{linenomath*}
The matrices $\Ud^+$ and $\Ud^-$ are formed as before, where the $\u = [\u_x \u_y]^T$.

\subsubsection{Flux-limiter implementation}
The advection matrix described above implements the first-order upwind flux that introduces numerical diffusion and smears the transported quantity significantly \citep{LeVeque1992}. To reduce numerical diffusion we implement a second-order Lax-Wendroff flux. Using the operators defined above, the advection matrix for the Lax-Wendroff flux is given by
\begin{linenomath*}
\begin{align}
    \A_{\mathrm{LxW}}(\u) = \A(\u) + \frac{1}{2} \Ud(\u) \left(\Ip - \Delta t \boldsymbol{\Delta}^{-1} \Ud(\u)\right) \boldsymbol{\Delta}\,\Gp,
\end{align}
\end{linenomath*}
Here $\Ud$ is a $\Nf\times\Nf$ diagonal matrix containing $\left|\u\right|$, where $\Nf = \Nfx + \Nfy$ is the total number of faces on the pressure grid.  The $\Nf\times\Nf$ diagonal matrix $\boldsymbol{\Delta}$ contains $\Nfx$ entries of $\Delta x$ followed by $\Nfy$ entries of $\Delta y$. The product $ \boldsymbol{\Delta}\, \Gp$ simply computes the change of the transported quantity rather than its gradient

The Lax-Wendroff flux introduced above introduces dispersive error that leads to oscillations in regions where the solution changes rapidly.  We introduce flux limiters to prevent these oscillations 
\begin{linenomath*}
\begin{align}
    \A_{\mathrm{FL}}(\u) = \A(\u) + \frac{1}{2} \Ud(\u) \left(\Ip - \Delta t \boldsymbol{\Delta}^{-1} \Ud(\u)\right) \boldsymbol{\Phi}_d(\boldsymbol{\theta}) \boldsymbol{\Delta}\, \Gp
\end{align}
\end{linenomath*}
where $\boldsymbol{\Phi}_d $ is an $\Nf\times\Nf$ diagonal matrix containing the flux limiter function $\phi$ for each face in the pressure grid. Here $\boldsymbol{\theta}$ is a $\Nf\times 1$ vector containing the smoothness indicator for every face. For details on the smoothness indicator and the different flux limiters available, please see \cite{LeVeque1992}. For all simulations presented here we have chosen the monotonized central (MC) limiter.  So that the final discrete advection diffusion operator is given by
\begin{linenomath*}
\begin{align}
    \L = \Dp\,\left(\A_{\mathrm{MC}}(\u) - \D_d(\H) \Gp\right).
\end{align}
\end{linenomath*}

\subsection{Verification}
\begin{figure}
    \centering
    \includegraphics[width = \textwidth]{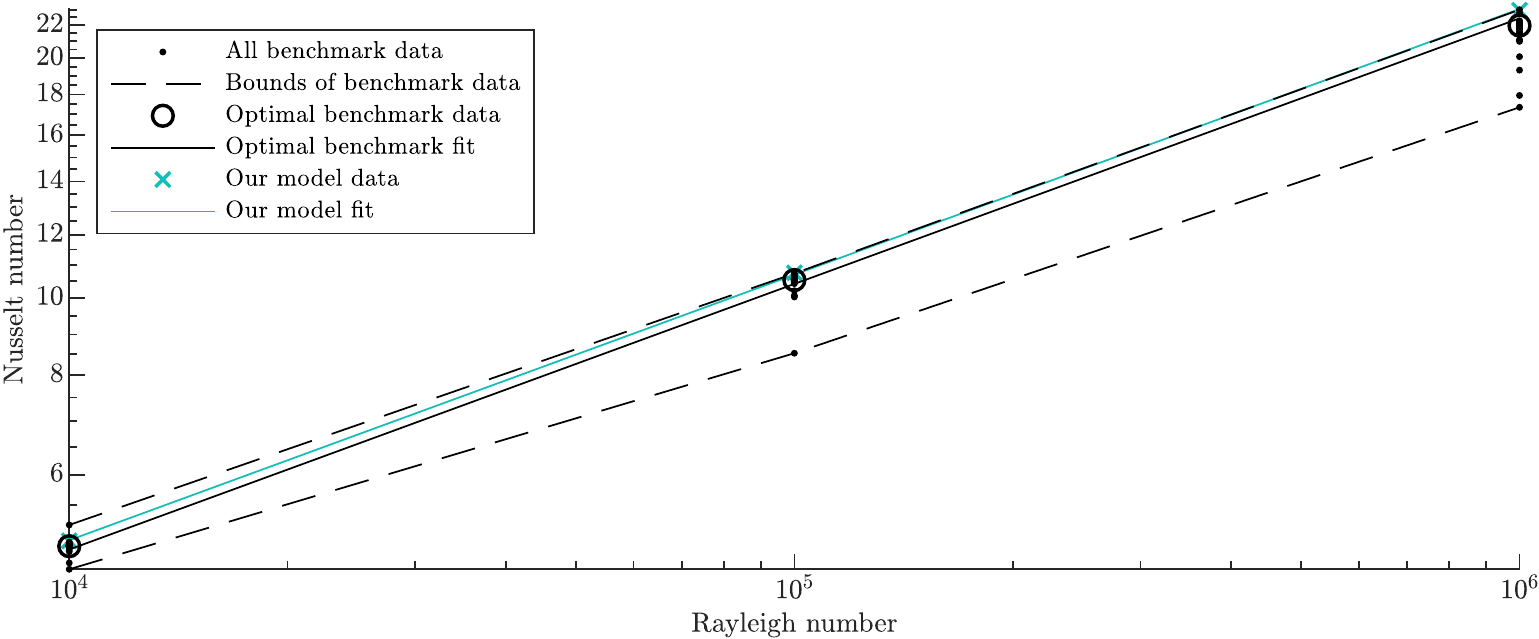}
    \caption{Rayleigh-Nusselt scaling for isoviscous convection. The output of the model presented here are shown compared to all benchmark data and the optimal benchmark values of \cite{Blankenbach1989}}
    \label{fig:nuRaScaling}
\end{figure}

\begin{figure}
    \centering
    \includegraphics[width = \textwidth]{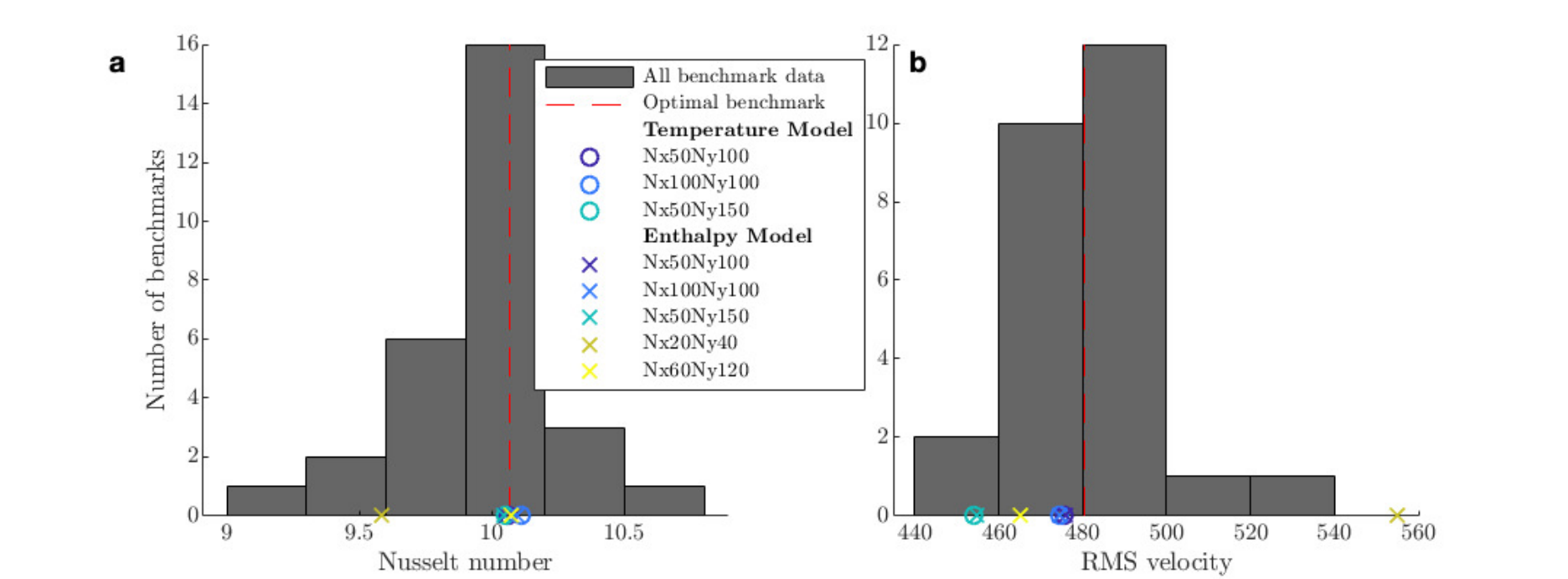}
    \caption{Temperature dependent viscosity benchmark values for (a) Nusselt number and (b) root mean squared velocity are compared to various grid resolutions of the model presented here \cite{Blankenbach1989}. Both the implementation of our model in the standard temperature formulation and the enthalpy formulation used in this paper are shown.}
    \label{fig:tempViscBench}
\end{figure}

We verify our numerical implementation against the standard convection benchmark of \citep{Blankenbach1989}. We benchmark both the isoviscous and temperature dependent viscosity case. With increasing grid resolution all Nusselt number benchmarks converge to the optimal benchmark provided. Our modeled values for the isoviscous case fall within 5\% of the optimal benchmark for Nusselt and average velocity at fine, 400 by 100, grid resolution at all Ra numbers. The Ra-Nu scaling is within the values presented by \cite{Blankenbach1989}, Figure \ref{fig:nuRaScaling}. The output from our temperature dependent viscosity benchmark falls within the range of possible benchmarks for both average velocity and Nusselt number, at grid resolutions near 100, Figure \ref{fig:tempViscBench}. Both the enthalpy and temperature formulated models give the same results for this benchmark, as they are physically identical models in this simulation (i.e.- no latent heat or changes in specific heat). We further check both the grid resolution we used for preliminary results, 40 by 20, and the ones presented in the paper, 120 by 60, the latter of which is well within the benchmark values for temperature dependent viscosity, and near the optimal value.

\begin{figure}
    \centering
    \includegraphics[width = \textwidth]{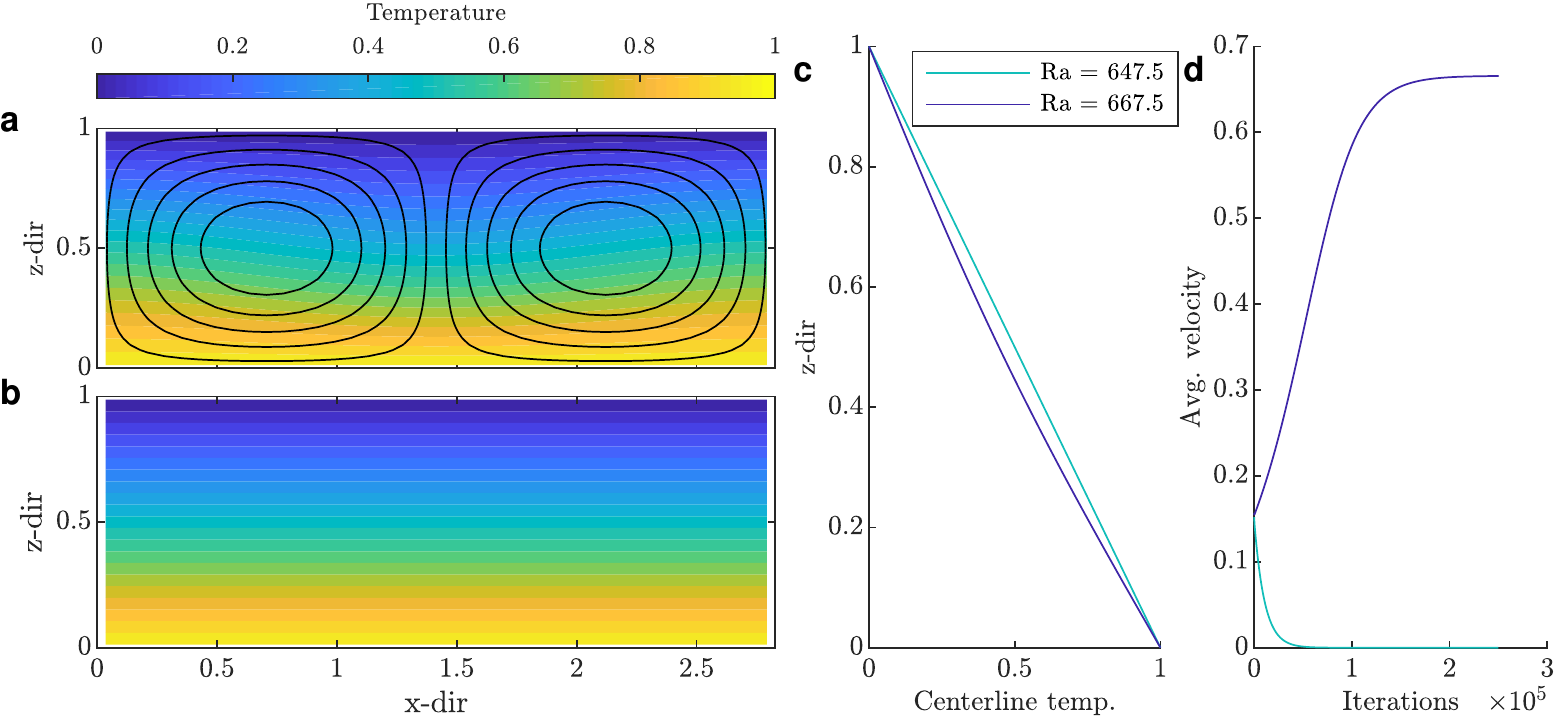}
    \caption{We test the stability 10 above (Ra = 667.5) and 10 below (Ra = 647.5) the analytically derived critical Rayleigh number (Ra = 657.5) for a 1 by $2\sqrt{2}$ box with a $2\sqrt{2}$ wavelength perturbation \citep{Turcotte2002}. We apply the same amplitude perturbation to both simulations. (a) The convective thermal profile at Ra = 667.5 and (b) the conductive thermal profile at Ra = 647.5. (c) The center temperature profile for both Ra number simulations. (d) The average velocity of both Ra number simulations.}
    \label{fig:stabilityBench}
\end{figure}

We further check the modeled onset of convection near the critical Ra number against analytical results \citep{Turcotte2002}. We check both 10 above and 10 below the isoviscous Ra critical, 657.5, with a wavelength perturbation of $2\sqrt{2}$, in exactly that wavelength domain with free slip boundaries. At Ra = 667.5 convection onsets, whereas at 647.5 the same perturbation decays to a purely conductive profile, Figure \ref{fig:stabilityBench}.

\section{Simulations}
 All simulations are started from a purely conductive state and we apply a $<0.175 K$ amplitude sinusoidal perturbation to onset convection. We use annual averages of the estimated surface temperature to stay consistent with the timescales of convection. We run our simulations on a domain of size $\left[0,\, \mathrm{Ar}\right]\times\left[0,\,1\right]$ with a 120 by 120 Cartesian grid. Increasing the aspect ratio beyond two has a negligible impact on convection \citep{Kalousova2017}. Streamlines are calculated as equally spaced contours of the streamfunction \citep{Batchelor2000}. We apply free slip boundary conditions on all sides, adiabatic side boundaries, and fix enthalpy at the base and surface of the domain. We calculate potential Ra combinations for Europa, Titan, and Enceladus by uniformly sampling the likely uncertainty in the parameter ranges for bottom viscosity, $10^{13} - 10^{15}$ \cite{Barr2009}, thickness (Titan 50-150 km, Europa 5-30 km, 10-50 km Enceladus; \cite{Vance2018}), and surface temperature (Europa 46-96 K, \cite{Ashkenazy2019}; Titan 89-94 K, \cite{Jennings2016}; Enceladus 48-63 K, \cite{Weller2019}). For Europa, we further estimate the bounds of non-dimensional tidal heating, $\Pi$, based off of thickness, tidal heating rate (2e-6 -- 8e-6 W/m$^3$) \citet{Tobie2003}, and surface temperature. For the response time comparison, the perturbation dies down before the energy within the ice shell increases. We align the onset of convection for both $c_p(T)$ and $c_p = c_p(T_b)$ simulations by beginning the comparison at epsilon (1e-4) increase in energy storage rate. We stop our response time simulations when the energy storage rate falls below 0.01.\\

%


\noindent\textbf{Data Set 1}
Excel file containing a summary of thermal conductivity data obtained from published works including tabulated data and data digitized from plots using the open source WebPlotDigitizer tool (\url{https://automeris.io/WebPlotDigitizer/citation.html}). The file also contains a summary of published thermal conductivity models describing the underlying data used in obtaining the fits, the valid temperature range, and the estimated uncertainty.






%
%


%
%
%
%
%

\bibliography{marc}

\clearpage


%
%
%
%

\begin{table*}[ht]
\footnotesize
\caption{Symbology, characteristic scales, and free parameters are given below. Characteristic scales are chosen at the base of the ice shell, i.e.- at the low pressure melting point of ice.} 
\centering 
\begin{tabular*}{\textwidth}{c @{\extracolsep{\fill}} l l l l} 
\hline  
Variable & Description & Value & Dimensions \\ [1ex] 
\hline 
$[\cdot]^\prime$ & superscript indicating dimensionless quantity &  [--] & [--] \\
$[\cdot]_c$ & subscript indicates characteristic quantity &  [--] & [--] \\
$\Delta[\cdot]$ & difference from bottom to top of ice shell &  [--] & [--] \\
$a,b,v$ & coefficients in general constitutive relationships & [--] & [--]\\
$c$ & coefficient in thermal conductivity relationship$^1$ & $1.97 \cdot 10^{-3}$ & W m$^{-1}$ K$^{-2}$\\
$\pi_1$, $\pi_2$, $\pi_3 \ $  & dimensionless intercept, slope, and exponent of constitutive relations & [--] & [--]\\
$k_c$ & characteristic thermal conductivity & $\sim 2.25$ & W m$^{-1}$ k$^{-1}$\\
$\Theta$ & the homologous temperature at the ice surface & [--] & [--]\\
$\eta_c$ & characteristic viscosity of ice$^2$ & [$10^{13}\ - 10^{15}$] & Pa s\\
$T_m$ , $T_b$ , $T_c$ & melting or basal temperature of ice shell & 273 & K\\
$T_s$ & surface temperature of ice shell; Europa$^3$, Titan$^4$, Enceladus$^5$ & $46-96$, $89-94$, $48-63$ & K\\
$E_a$ & ice viscosity activation energy$^6$ & 50,000 & J mol$^{-1}$\\
$R$ & universal gas constant & 8.314 & J K$^{-1}$ mol$^{-1}$\\
$A$ & exponent in viscosity law & $\frac{\mathrm{E}_a}{\mathrm{R\ T}_m}$ & [--]\\
$x_c$ , $d$ & thickness of ice shell$^7$; Europa, Titan, Enceladus & $5-30$, $50-150$, $10-50$ & km\\ 
$w$ & width of domain & $2d$ & km\\
$\rho_c$ & characteristic density & 917 & kg m$^{-3}$\\
$c_{p,c}$ & characteristic specific heat$^8$ & 2106.1 & J kg$^{-1}$ K$^{-1}$\\
$H_c$ & characteristic volumetric enthalpy & $\rho_c c_{p,c} \Delta T$ & J m$^{-3}$\\
$\kappa_c$ & characteristic thermal diffusivity & $\frac{k_c}{c_{p,c} \rho_c}$ & m$^2$ s$^{-1}$\\
$t_c$ & diffusive time scale & $\frac{x_c^2}{\kappa_c}$ & s\\
$u_c$ & advective velocity scale & $\frac{x_c}{t_c}$ & m s$^{-1}$\\
$P_c$ & characteristic pressure & $\frac{\eta_c k_c}{x_c^2}$ & Pa\\
$g$ & gravity; Europa, Titan, Enceladus & $1.315, \ 1.352,\ 0.111$ & m s$^{-2}$ \\
$G_c$, $G_{max}$ & characteristic tidal heating rate; Europa$^9$ & $2\cdot 10^{-6} - 8\cdot 10^{-6}$ & W m$^{-3}$\\
$\hat{z}$ & vertical spatial coordinate vector &  [--] & [--] \\
Ra & Rayleigh number & $\frac{g x_c^3 \Delta \rho}{\eta_c k_c}$ & [--]\\
$\tidal$ & Dimensionless tidal heating rate & $\frac{G_c x_c^2}{k_c \Delta T}$ & [--]\\
$Ar$ & Aspect ratio & $d/w$ & [--]\\
[1ex]
\hline 
\end{tabular*}
$^1$\citep{Andersson2005}, $^2$\citep{Barr2009}, $^3$\citep{Jennings2016}, $^4$\citep{Ashkenazy2019}, $^5$\citep{Weller2019}, $^6$\citep{Goldsby2001}, $^7$\citep{Vance2018}, $^8$\citep{Ellsworth1983}, $^9$\citep{Tobie2003}
\label{tab:SymChar} 
\end{table*}

\begin{table}
\footnotesize
\caption{Dimensional and dimensionless coefficients for the temperature-dependent material properties.}\label{tab:prop_coeff}
\centering 
\begin{tabular}{|c|c|c|c|c|c|c|c|l}
\hline
& \multicolumn{3}{c}{dimensional} & \multicolumn{3}{c}{dimensionless}&\\ \hline
 property & $a$ & $b$ & $v$ & $\pi_1$ &$\pi_2$ & $\pi_3$ & reference \\ \hline
$k$ & 2.26 & 0 & 1 & 1 & 0 & 0 & (constant)\\ \hline
$k$ & 0.4685 & 488.12 & -1 & $\frac{a}{a+b/T_m}$ & $\frac{b/T_m}{a+b/T_m}$ & -1 & \citep{Hobbs1974} \\ \hline
$k$ & 0 & 2135 & -1.235 & 0 & 1 & $v$ & \citep{Rabin2000} \\ \hline
$k$ & 0 & 612 & -1 & 0 & 1 & -1 & (this work) \\ \hline
$c_p$  & 185 & 7.037 & 1 & $\frac{a}{a+bT_m}$  & $\frac{bT_m}{a+bT_m}$ & 1& \citep{Ellsworth1983} \\ \hline
$\eta$  & 6013.95 & 22.029 & -1  & $\frac{a}{a+b/T_m}$ & $\frac{b/T_m}{a+b/T_m}$  & -1 & \citep{Goldsby2001} \\ \hline
\end{tabular}
\end{table}